\documentclass[longauth,letter]{aa}

\usepackage{graphicx}
\usepackage{amsmath}
\usepackage{longtable}
\usepackage[varg]{txfonts}
\usepackage[citecolor=blue,colorlinks=true]{hyperref}

\begin{document}

\title{Transit timing variations of AU\,Microscopii b and c}

\titlerunning{TTVs of AU\,Mic\,b and c}

\author{
Gy. M. Szabó$^{1,2}$, 
Z. Garai$^{1,2,3}$, 
A. Brandeker$^{4}$, 
D. Gandolfi$^{5}$, 
T. G. Wilson$^{6}$, 
A. Deline$^{7}$, 
G. Olofsson$^{4}$, 
A. Fortier$^{8,9}$, 
D. Queloz$^{7,10}$, 
L. Borsato$^{11}$, 
F. Kiefer$^{12}$, 
A. Lecavelier des Etangs$^{12}$, 
M. Lendl$^{7}$, 
L. M. Serrano$^{5}$, 
S. Sulis$^{13}$, 
S. Ulmer Moll$^{7}$, 
V. Van Grootel$^{14}$, 
Y. Alibert$^{8}$, 
R. Alonso$^{15,16}$, 
G. Anglada$^{17,18}$, 
T. Bárczy$^{19}$, 
D. Barrado y Navascues$^{20}$, 
S. C. C. Barros$^{21,22}$, 
W. Baumjohann$^{23}$, 
M. Beck$^{7}$, 
T. Beck$^{8}$, 
W. Benz$^{8,9}$, 
N. Billot$^{7}$, 
A. Bonfanti$^{23}$, 
X. Bonfils$^{24}$, 
C. Broeg$^{8,9}$, 
J. Cabrera$^{25}$, 
S. Charnoz$^{26}$, 
A. Collier Cameron$^{6}$, 
Sz. Csizmadia$^{25}$, 
M. B. Davies$^{27}$, 
M. Deleuil$^{13}$, 
L. Delrez$^{14,28}$, 
O. Demangeon$^{21,22}$, 
B.-O. Demory$^{9}$, 
D. Ehrenreich$^{7}$, 
A. Erikson$^{25}$, 
L. Fossati$^{23}$, 
M. Fridlund$^{29,30}$, 
M. Gillon$^{28}$, 
M. Güdel$^{31}$, 
K. Heng$^{9,32}$, 
S. Hoyer$^{13}$, 
K. G. Isaak$^{33}$, 
L. L. Kiss$^{34,48}$, 
J. Laskar$^{35}$, 
C. Lovis$^{7}$, 
D. Magrin$^{11}$, 
P. F. L. Maxted$^{36}$, 
M. Mecina$^{37}$, 
V. Nascimbeni$^{11}$, 
R. Ottensamer$^{37}$, 
I. Pagano$^{39}$, 
E. Pallé$^{15}$, 
G. Peter$^{40}$, 
G. Piotto$^{11,41}$, 
D. Pollacco$^{32}$, 
R. Ragazzoni$^{11,41}$, 
N. Rando$^{42}$, 
H. Rauer$^{25,43,44}$, 
I. Ribas$^{17,18}$, 
N. C. Santos$^{21,22}$, 
M. Sarajlic$^{8}$, 
G. Scandariato$^{39}$, 
D. Ségransan$^{7}$, 
A. E. Simon$^{8}$, 
A. M. S. Smith$^{25}$, 
S. Sousa$^{21}$, 
M. Steller$^{23}$, 
N. Thomas$^{8}$, 
S. Udry$^{7}$, 
F. Verrecchia$^{45,46}$, 
N. Walton$^{47}$,
D. Wolter$^{25}$
}

\authorrunning{Szab\'o et al.}

\institute{
$^{1}$ ELTE Eötvös Loránd University, Gothard Astrophysical Observatory, 9700 Szombathely, Szent Imre h. u. 112, Hungary\\
$^{2}$ MTA-ELTE Exoplanet Research Group, 9700 Szombathely, Szent Imre h. u. 112, Hungary\\
$^{3}$ Astronomical Institute, Slovak Academy of Sciences, 05960 Tatranská
Lomnica, Slovakia\\
$^{4}$ Department of Astronomy, Stockholm University, AlbaNova University Center, 10691 Stockholm, Sweden\\
$^{5}$ Dipartimento di Fisica, Universita degli Studi di Torino, via Pietro Giuria 1, I-10125, Torino, Italy\\
$^{6}$ Centre for Exoplanet Science, SUPA School of Physics and Astronomy, University of St Andrews, North Haugh, St Andrews KY16 9SS, UK\\
$^{7}$ Observatoire Astronomique de l'Université de Genève, Chemin Pegasi 51, Versoix, Switzerland\\
$^{8}$ Physikalisches Institut, University of Bern, Gesellsschaftstrasse 6, 3012 Bern, Switzerland\\
$^{9}$ Center for Space and Habitability, Gesellsschaftstrasse 6, 3012 Bern, Switzerland\\
$^{10}$ Cavendish Laboratory, JJ Thomson Avenue, Cambridge CB3 0HE, UK\\
$^{11}$ INAF, Osservatorio Astronomico di Padova, Vicolo dell'Osservatorio 5, 35122 Padova, Italy\\
$^{12}$ Institut d'astrophysique de Paris, UMR7095 CNRS, Université Pierre \& Marie Curie, 98bis blvd. Arago, 75014 Paris, France\\
$^{13}$ Aix Marseille Univ, CNRS, CNES, LAM, 38 rue Frédéric Joliot-Curie, 13388 Marseille, France\\
$^{14}$ Space sciences, Technologies and Astrophysics Research (STAR) Institute, Universit{\'e} de Liège, Allée du 6 Août 19C, 4000 Liège, Belgium\\
$^{15}$ Instituto de Astrofisica de Canarias, 38200 La Laguna, Tenerife, Spain\\
$^{16}$ Departamento de Astrofisica, Universidad de La Laguna, 38206 La Laguna, Tenerife, Spain\\
$^{17}$ Institut de Ciencies de l'Espai (ICE, CSIC), Campus UAB, Can Magrans s/n, 08193 Bellaterra, Spain\\
$^{18}$ Institut d'Estudis Espacials de Catalunya (IEEC), 08034 Barcelona, Spain\\
$^{19}$ Admatis, 5. Kandó Kálmán Street, 3534 Miskolc, Hungary\\
$^{20}$ Depto. de Astrofisica, Centro de Astrobiologia (CSIC-INTA), ESAC campus, 28692 Villanueva de la Cañada (Madrid), Spain\\
$^{21}$ Instituto de Astrofisica e Ciencias do Espaco, Universidade do Porto, CAUP, Rua das Estrelas, 4150-762 Porto, Portugal\\
$^{22}$ Departamento de Fisica e Astronomia, Faculdade de Ciencias, Universidade do Porto, Rua do Campo Alegre, 4169-007 Porto, Portugal\\
$^{23}$ Space Research Institute, Austrian Academy of Sciences, Schmiedlstrasse 6, A-8042 Graz, Austria\\
$^{24}$ Université Grenoble Alpes, CNRS, IPAG, 38000 Grenoble, France\\
$^{25}$ Institute of Planetary Research, German Aerospace Center (DLR), Rutherfordstrasse 2, 12489 Berlin, Germany\\
$^{26}$ Université de Paris, Institut de physique du globe de Paris, CNRS, F-75005 Paris, France\\
$^{27}$ Centre for Mathematical Sciences, Lund University, Box 118, SE 221 00, Lund, Sweden\\
$^{28}$ Astrobiology Research Unit, Université de Liège, Allée du 6 Août 19C, B-4000 Liège, Belgium\\
$^{29}$ Leiden Observatory, University of Leiden, PO Box 9513, 2300 RA Leiden, The Netherlands\\
$^{30}$ Department of Space, Earth and Environment, Chalmers University of Technology, Onsala Space Observatory, 43992 Onsala, Sweden\\
$^{31}$ University of Vienna, Department of Astrophysics, Türkenschanzstrasse 17, 1180 Vienna, Austria\\
$^{32}$ Department of Physics, University of Warwick, Gibbet Hill Road, Coventry CV4 7AL, United Kingdom\\
$^{33}$ Science and Operations Department - Science Division (SCI-SC), Directorate of Science, European Space Agency (ESA), European Space Research and Technology Centre (ESTEC),
Keplerlaan 1, 2201-AZ Noordwijk, The Netherlands\\
$^{34}$ Konkoly Observatory, Research Centre for Astronomy and Earth Sciences, 1121 Budapest, Konkoly Thege Miklós út 15-17, Hungary\\
$^{35}$ IMCCE, UMR8028 CNRS, Observatoire de Paris, PSL Univ., Sorbonne Univ., 77 av. Denfert-Rochereau, 75014 Paris, France\\
$^{36}$ Astrophysics Group, Keele University, Staffordshire, ST5 5BG, United Kingdom\\
$^{37}$ Department of Astrophysics, University of Vienna, Tuerkenschanzstrasse 17, 1180 Vienna, Austria\\
$^{38}$ {MTA-ELTE Lend{\"u}let "Momentum" Milky Way Research Group, Hungary}\\
$^{39}$ INAF, Osservatorio Astrofisico di Catania, Via S. Sofia 78, 95123 Catania, Italy\\
$^{40}$ Institute of Optical Sensor Systems, German Aerospace Center (DLR), Rutherfordstrasse 2, 12489 Berlin, Germany\\
$^{41}$ Dipartimento di Fisica e Astronomia "Galileo Galilei", Universita degli Studi di Padova, Vicolo dell'Osservatorio 3, 35122 Padova, Italy\\
$^{42}$ ESTEC, European Space Agency, 2201AZ, Noordwijk, NL\\
$^{43}$ Center for Astronomy and Astrophysics, Technical University Berlin, Hardenberstrasse 36, 10623 Berlin, Germany\\
$^{44}$ Institut für Geologische Wissenschaften, Freie UniversitÃ¤t Berlin, 12249 Berlin, Germany\\
$^{45}$ Space Science Data Center, ASI, via del Politecnico snc, 00133 Roma, Italy\\
$^{46}$ INAF, Osservatorio Astronomico di Roma, via Frascati 33, 00078 Monte Porzio Catone (RM), Italy \\
$^{47}$ Institute of Astronomy, University of Cambridge, Madingley Road, Cambridge, CB3 0HA, United Kingdom\\
$^{48}$ ELTE E\"otv\"os Lor\'and University, Institute of Physics, P\'azm\'any P\'eter s\'et\'any 1/A, 1117 Budapest, Hungary
}

\date{Received date / Accepted date }

\abstract{
Here we report large-amplitude transit timing variations (TTVs) for AU\,Microcopii b and c as detected in combined \textit{TESS} (2018, 2020) and \textit{CHEOPS} (2020, 2021) transit observations.
AU\,Mic is a young planetary system with a debris disk and two transiting warm Neptunes. A TTV on the order of several minutes was previously reported for AU\,Mic b, which was suggested to be an outcome of mutual perturbations between the planets in the system. In 2021, we observed AU\,Mic b (five transits) and c (three transits) with the \textit{CHEOPS} space telescope to follow-up the TTV of AU\,Mic b and possibly detect a TTV for AU\,Mic c. When analyzing \textit{TESS} and \textit{CHEOPS} 2020--2021 measurements together, we find that a prominent TTV emerges with a full span of $\geq23$ minutes between the two TTV extrema. Assuming that the period change results from a periodic process ---such as mutual perturbations--- we demonstrate that the times of transits in the summer of 2022 are expected to be 30--85 minutes later than predicted by the available linear ephemeris.
}

%\begin{keywords}
%planetary systems: individual: AU Mic
%\end{keywords}

\maketitle

\section{Introduction}

AU~Microscopii is the epitome of a young planetary system, where planets orbit a late-type star, with possible star--planet interactions.
Its age is estimated to be 22\,Myr \citep{2014MNRAS.445.2169M}, and as a $\beta$~Pictoris Moving Group \citep{2006A&A...460..695T} member, it has a dynamical trace-back age of $18.5\pm2$~Myr \citep{2020A&A...642A.179M}, making it\ one of the youngest known exoplanet systems. AU\,Mic hosts
two transiting warm Neptunes near mean-motion resonances \citep{2020Natur.582..497P}.
Recurrent spot occultations along the transit chord are observed thanks to a 7:4 spin--orbit commensurability between the orbital period of AU\,Mic\,b and the stellar rotation \citep{2021A&A...654A.159S}.
%The host star is very active \citep{2021arXiv210903924G}
%and a debris disc of a complex structure 
%is also present in the system
%\citep{2019ApJ...883L...8W}.

Eight transits of AU\,Mic\,b have been published so far.
\textit{TESS} \citep{2014SPIE.9143E..20R} observed two and three transits in 2018 and 2020, respectively \citep{2021arXiv210903924G}, and \textit{CHEOPS} \citep{2021ExA....51..109B} observed three transits in 2020 \citep[previously reported in][]{2021A&A...654A.159S}. All published data on AU\,Mic c so far come from \textit{TESS}, covering one transit in 2018 and two in 2020 \citep{2020Natur.582..497P,2021arXiv210903924G}. The previous observations led to somewhat inconsistent period estimates for AU\,Mic\,b (with a scatter significantly larger than expected from the estimated timing uncertainty), which has recently been suggested to reflect TTVs on the order of 80\,s \citep[reported by][]{2021arXiv210903924G} or 3\,min  \citep[by][]{2021A&A...649A.177M}.

Here we present new photometric observations of AU\,Mic~b and AU\,Mic c carried out with the \textit{CHEOPS} space telescope from July through September, 2021. We describe the observations and data-processing methods in Sect.~2 and present the results in Sect.~3.

\section{Observations and data processing}

\begin{table*}
    \centering
    \caption{Logs of AU Mic observations by \textit{CHEOPS} included in this Letter. The time notation follows the ISO-8601 convention. The File Key supports the fast identification of the observations in the \textit{CHEOPS} archive.}
    \begin{tabular}{cccccccc}
    \hline
    \hline
    \noalign{\smallskip}
    Visit ID & Start Date & End Date & File Key & \textit{CHEOPS} & Integ. & Co-added & Num. of     \\
      & (2021) & (2021) & & product & time (s) & exposures & frames   \\
    \noalign{\smallskip}
    \hline
    \noalign{\smallskip}
%    4 / AU Mic c \#1 & {\small 07-21 11:09:57} & {\small 07-21 23:52:03} & {\small PR100010\_TG003201} & Subarray & 42 & 3\,s $\times$ 14 & 743 \\
%        & & & & {\it Imagettes} & 3 & --- & 10\,402 \\
    AU Mic b 21-07-26 & {\small 07-26 11:27:13} & {\small 07-26 22:34:04} & {\small PR100010\_TG003001} & Subarray & 42 & 3\,s $\times$ 14 & 669 \\
        & & & & {\it Imagettes} & 3 & --- & 9366 \\
    AU Mic c 21-08-09 & {\small 08-09 04:59:15} & {\small 08-09 19:37:47} & {\small PR100010\_TG003401} & Subarray & 42 & 3\,s $\times$ 14 & 1029 \\
        & & & & {\it Imagettes} & 3 & --- & 14\,406 \\
    AU Mic b 21-08-12 & {\small 08-12 08:25:41} & {\small 08-12 19:53:00} & {\small PR100010\_TG003601} & Subarray & 42 & 3\,s $\times$ 14 & 839 \\
        & & & & {\it Imagettes} & 3 & --- & 11\,746 \\
    AU Mic c 21-08-28 & {\small 08-28 02:09:13} & {\small 08-28 16:35:03} & {\small PR100010\_TG003402} & Subarray & 42 & 3\,s $\times$ 14 & 907 \\
        & & & & {\it Imagettes} & 3 & --- & 12\,698 \\
    AU Mic b 21-08-29 & {\small 08-29 05:17:41} & {\small 08-29 16:44:59} & {\small PR100010\_TG003701} & Subarray & 42 & 3\,s $\times$ 14 & 667 \\
        & & & & {\it Imagettes} & 3 & --- & 9338 \\
    AU Mic b 21-09-06 & {\small 09-06 17:38:41} & {\small 09-07 05:05:59} & {\small PR100010\_TG003101} & Subarray & 42 & 3\,s $\times$ 14 & 643 \\
        & & & & {\it Imagettes} & 3 & --- & 9002 \\
%    11 / AU Mic b \#8 & {\small 09-23 13:15:13} & {\small 09-24 00:42:31} & {\small PR100010\_TG003901} & Subarray & 42 & 3\,s $\times$ 14 & 519 \\
%        & & & & {\it Imagettes} & 3 & --- & 7266 \\
    \noalign{\smallskip}
  \hline
    \end{tabular}
    \label{tab:cheops_log}
\end{table*}

During the 2021 opposition, we observed five transits of AU\,Mic\,b and three transits of AU\,Mic\,c. Four of the five AU\,Mic\,b transits and two of the AU\,Mic\,c transits are appropriate for transit timing analysis. For the fifth AU\,Mic\,b transit (21-09-25), both the ingress and egress are missing because of gaps in the data. These transits will be analyzed in a forthcoming publication. Similarly to the third 2020 observation with \textit{CHEOPS}, we used short exposure times of 3\,s to better resolve possible flares. The brightness of the star \citep[V = 8.6\,mag and GAIA G = 7.843\,mag;][]{2012AcA....62...67K,2018A&A...616A...1G} ensured an adequate signal despite the short exposures. See Table~\ref{tab:cheops_log} for the observations log.
In the \textit{CHEOPS} Proposal Handling Tool, we set up observation windows { with an observation length} covering seven (AU\,Mic\,b) and nine  \textit{CHEOPS} orbits (AU\,Mic\,c), with one \textit{CHEOPS} orbit lasting 98.77\,min. We centered each visit at the predicted mid-transit time and observed for the entire transit duration, adding at least 1.5 CHEOPS orbits  on each side in order to have a reasonably long out-of-transit baseline. The efficiency of the observations varied between 55\,\% and 90\,\%.

The sub-array frames were automatically processed with the \textit{CHEOPS} Data Reduction Pipeline \citep[DRP;][]{2020A&A...635A..24H}. In addition to the sub-arrays, there are \textit{imagettes} available for each exposure. The imagettes are images of 30 pixels in radius centered on the target and do not need to be co-added before download owing to their smaller size. We used a tool specifically developed for photometric extraction of imagettes using point-spread function (PSF) photometry, PIPE (\textit{PSF imagette photometric extraction}; for more details of how it was applied to AU\,Mic data, we refer to \citealt{2021A&A...654A.159S}). 
The PIPE photometry  has a signal-to-noise ratio (S/N) comparable to that of DRP photometry, but has a lower cadence, allowing better identification of flares \citep{2021A&A...654A.159S}.

In this Letter, we analyze the PIPE reduction averaged to 15~s cadence for better S/N of individual points.
The data reduced with both DRP and PIPE  will be available on VizieR after the publication of this Letter.

\subsection{Pre-processing of the light curves}

Due to the rotation of the host star, the stellar brightness slowly varied during the observations. This was removed by fitting a fourth-order polynomial to the out-of-transit light curve segments before starting the analysis. 
%The light curve solving algorithm included a Gaussian Process (GP) module that was able to remove the effects of slow brightness variations during the transit. 

Because flares can severely bias the transit model, we masked them out during the transits before the modeling. The longest flare with at least three major, complex humps occurred around the ingress phase of the \textit{CHEOPS} 21-07-26 visit, and only the egress of this transit could be kept for further analysis. A shorter lasting flare was observed during the transits of the \textit{CHEOPS} 21-08-29 and \textit{CHEOPS} 21-09-23 visits, which nevertheless did not affect the ingress and egress phases.

We show the raw \textit{CHEOPS} light curves after a subtraction of the polynomial slow-trend model and the image synthesis model of background contamination in Fig.~\ref{fig:rawLightCurves}. We note that the segments of light curves contaminated by flares (in smaller dots in the figure) were omitted from the fit. 

\begin{figure}
    \centering
    \includegraphics[viewport=12 120 440 380, width=\columnwidth]{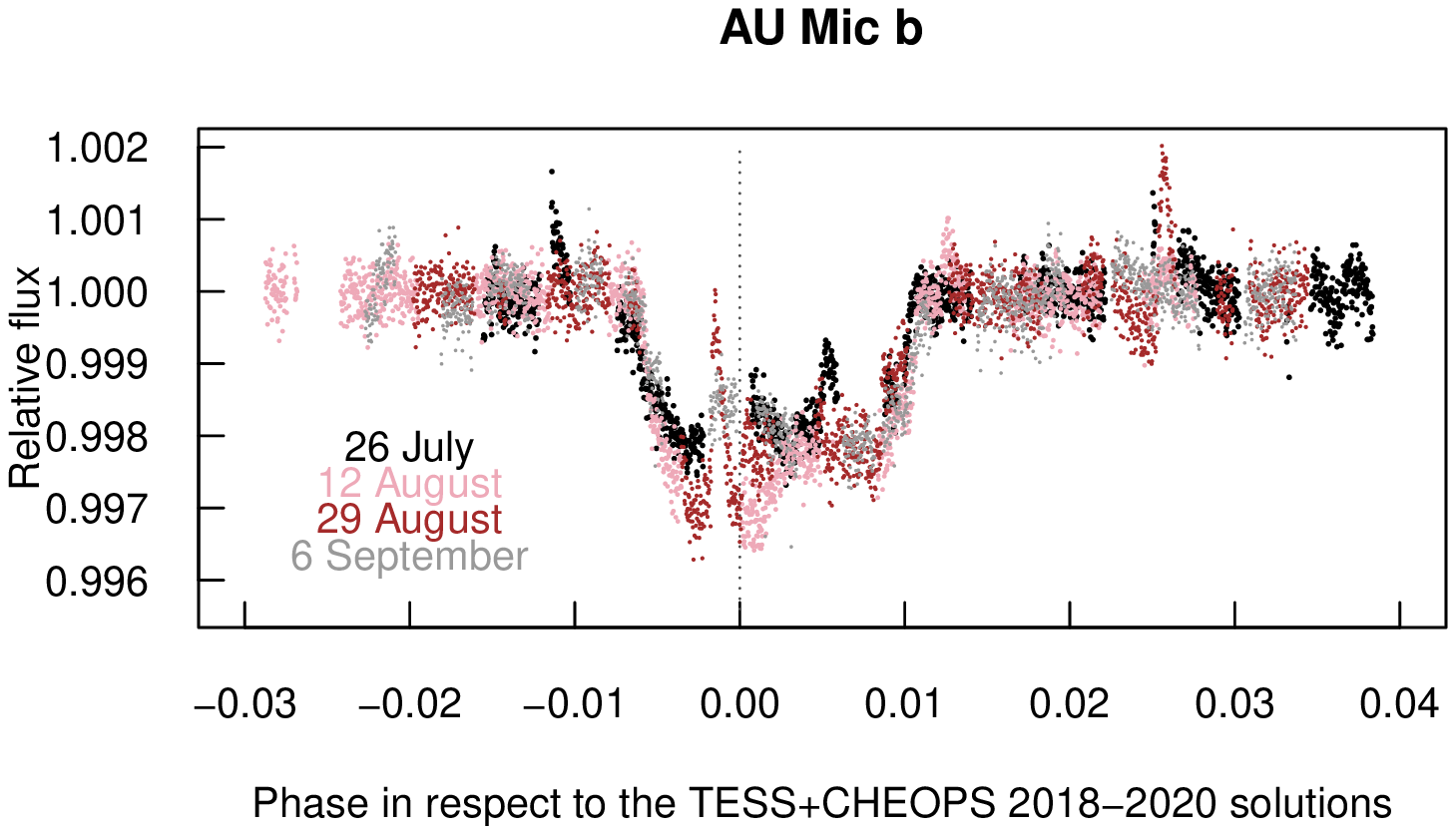}
    \includegraphics[viewport=12 120 440 380, width=\columnwidth]{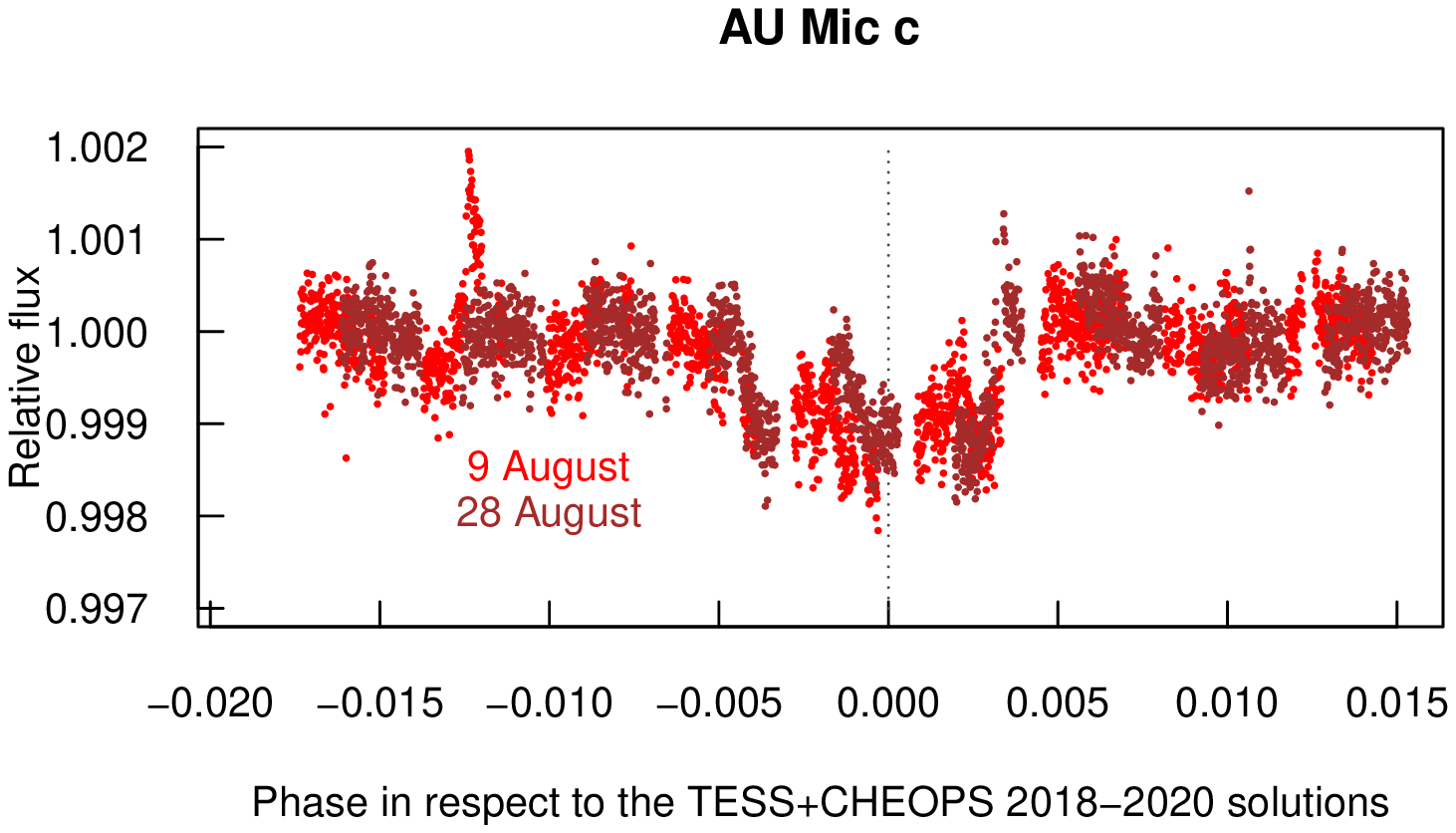}
    \caption{Phased transit light curves of AU Mic b (upper panel) and AU Mic c (lower panel). The ordinate shows the orbital phase using the ephemeris of Szab\'o et al. (2021): $T_c=2\,459\,041.28272$, $P=8.462995$ in the case of AU\,Mic\,b, and $T_c=2\,458\,342.2223$, $P=18.859019$ in the case of AU\,Mic\,c. We highlight the huge phase shift of the 2021 transits with respect to the earlier linear ephemeris based on \textit{TESS} and \textit{CHEOPS} data from 2018 to 2020.}
    \label{fig:lcs}
\end{figure}

A visual inspection of the light curve led us to conclude that the level of red noise is not negligible. 
A quantitative analysis showed that the residuals after fitting the transits on the 2021 light curves are on average 280\,ppm, higher than the 125\,ppm residuals we found in the 2020 data (see also Fig.~\ref{fig:lcs} and its discussion). The increasing red noise can be attributed to an increased stellar activity in 2021 compared to 2020. 
The increasing activity between 2018 and 2020 was also observed by \textit{TESS} \citep{2021arXiv210903924G,2021A&A...649A.177M}. The adverse effects of increased activity are stronger for AU\,Mic\,c transits because of their shallowness, making the transit parameters more sensitive to red noise.  The changing spot coverage of the star leads to a bias in the planet size ($R_p/R_\star$) parameter as well (see Tables~\ref{tab:solutionsb} and \ref{tab:solutionsc} and the discussions that follow them). 

Because of the spin--orbit commensurability \citep{2021A&A...654A.159S} of AU\,Mic\,b, the transits 21-07-26 and 21-08-29 are observed in front of the same stellar longitude. This longitude also coincides with the \textit{CHEOPS} 20-08-21 and \textit{CHEOPS} 20-09-24 observations shown in \cite{2021A&A...654A.159S}, but the change in the spot map does not allow a direct comparison between the years.

The phased transit light curves of both AU\,Mic\,b and AU\,Mic\,c are shown in Fig.~\ref{fig:lcs}. 
In both panels of this figure, the period $P$ and epoch $T_c$ of the transit times have been adopted from publications based on 2018--2020 data
\citep[AU\,Mic\,b and c:][respectively]{2021A&A...649A.177M,2021arXiv210903924G}
to reflect the dramatic shift of the transits in reference to these linear ephemeris. The mid-transit of AU\,Mic\,b is shifted toward the positive phase coordinates, and a slight shift toward negative values is suspected in the case of AU\,Mic\,c. This is an impressive representation of how much the behavior of both planets changed within less than a year.

Both panels of Fig.~\ref{fig:lcs} show the \textit{CHEOPS} datasets without masking out the flares. In the case of AU\,Mic\,b, we see prominent anomalies mostly during the ingress phase and the start of the transit floor (around 0.00 phase coordinate according to the phase definition in Fig.~\ref{fig:lcs}). The ingress phase of the 21-07-26 transit is shallower than the other ones, which is further evidence of a positive anomaly at around phase 0.008 which is due to a small flare. The 21-08-12 transit is significantly steeper than the other ones, which is likely the result of a different spot distribution (e.g., the beginning of the transit chord not covered by spots). 

The 21-08-12 and 21-08-29 transits are exactly 3.5 stellar rotations apart and are therefore observed at opposite stellar longitudes. Interestingly, both of these transits have a ``brightening'' near the center of the transit, as if both of these transits happened in front of a spot. However, the presence of two large spots on opposite sides of the star is compatible with the rotation light curve of the star showing two minima during one rotation \citep{2020Natur.582..497P,2021A&A...649A.177M,2021A&A...654A.159S,2021arXiv210903924G}. 

The phased transit curve of AU\,Mic\,c shows an unlucky coverage: the egress phase is within the data gaps in the case of both observations. There are also many residuals during the transits that are likely due to the presence of spot occultations. There is little resemblance between the residuals of the two planets. However, the impact parameter $b$ is known to be different for the planets \citep{2021A&A...649A.177M,2021arXiv210903924G}, which means that the two transit chords map to different parts of the star. The observed residuals are consistent with this interpretation.

\subsection{The transit model}

\begin{figure}
    \centering
    \includegraphics[bb=16 116 440 400,width=0.9\columnwidth]{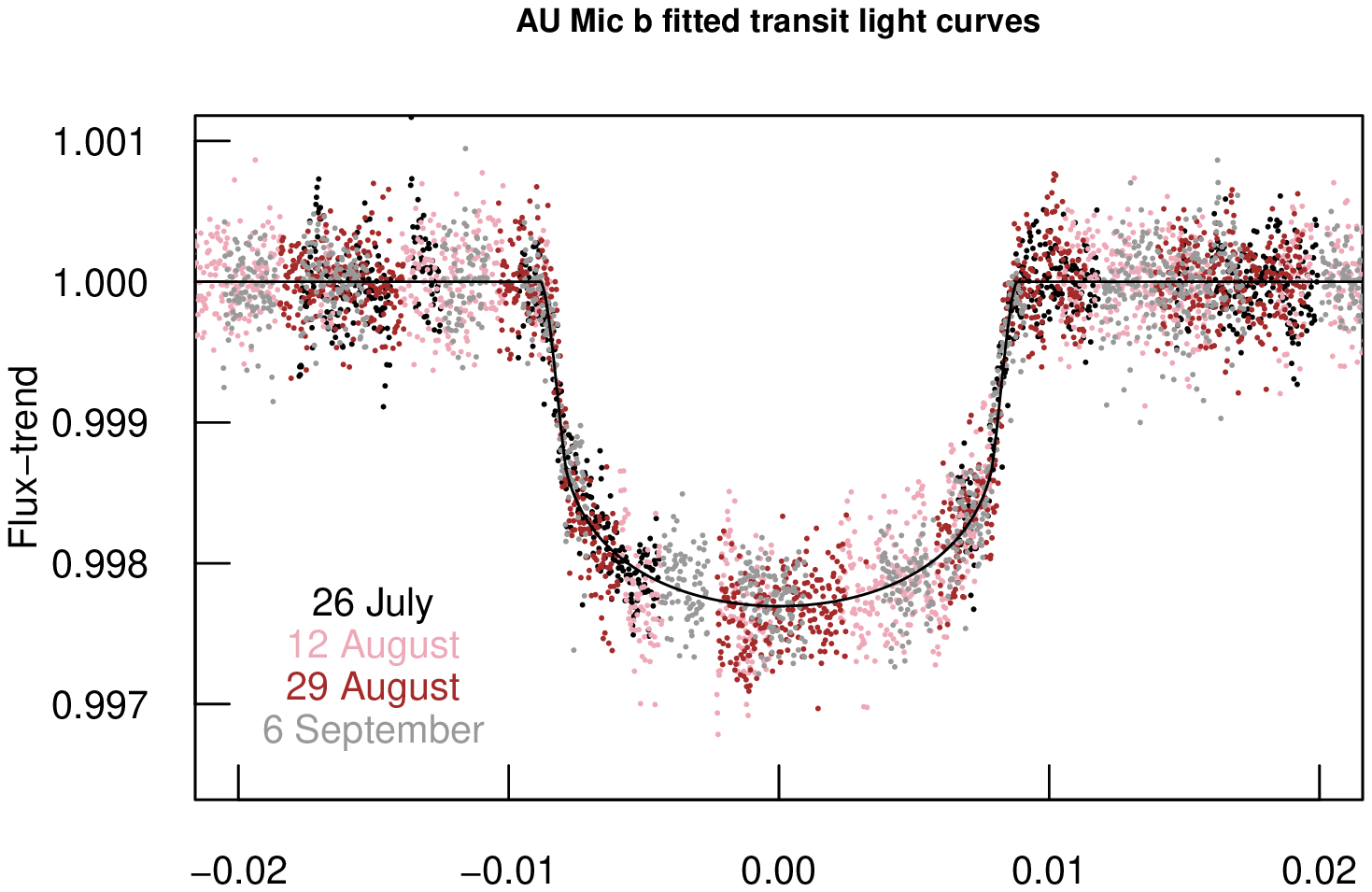}
    \includegraphics[bb=16 116 440 330, width=0.9\columnwidth]{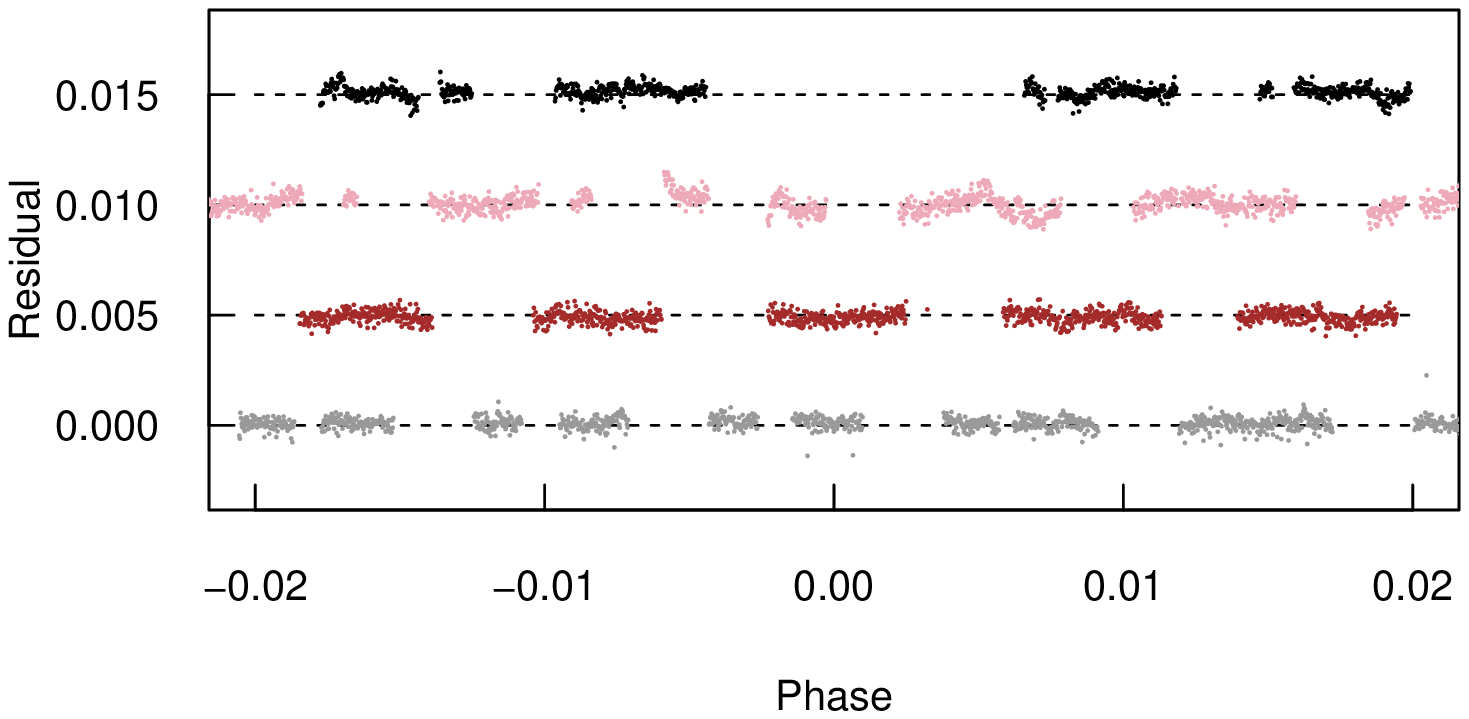}
    \includegraphics[bb=16 150 440 370,width=0.9\columnwidth]{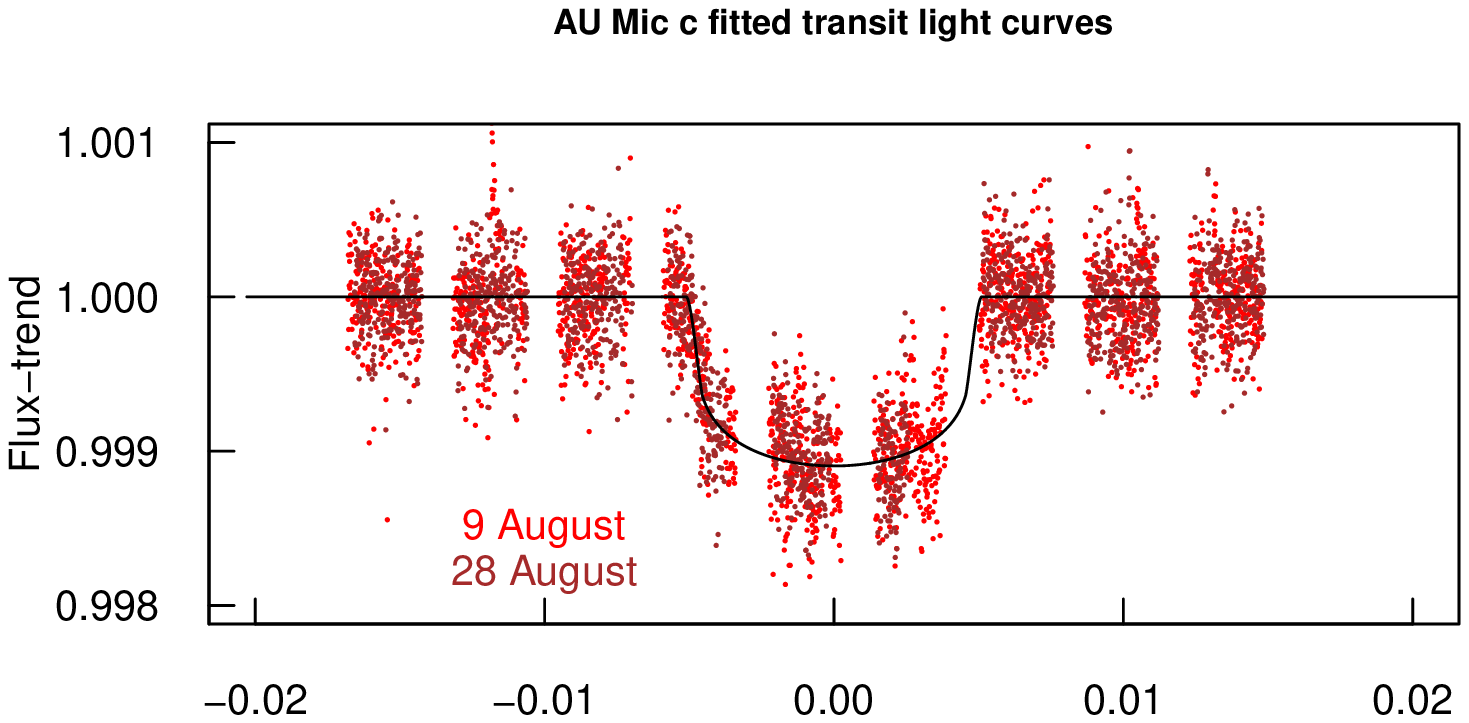}
    \includegraphics[bb=16 116 440 300, width=0.9\columnwidth]{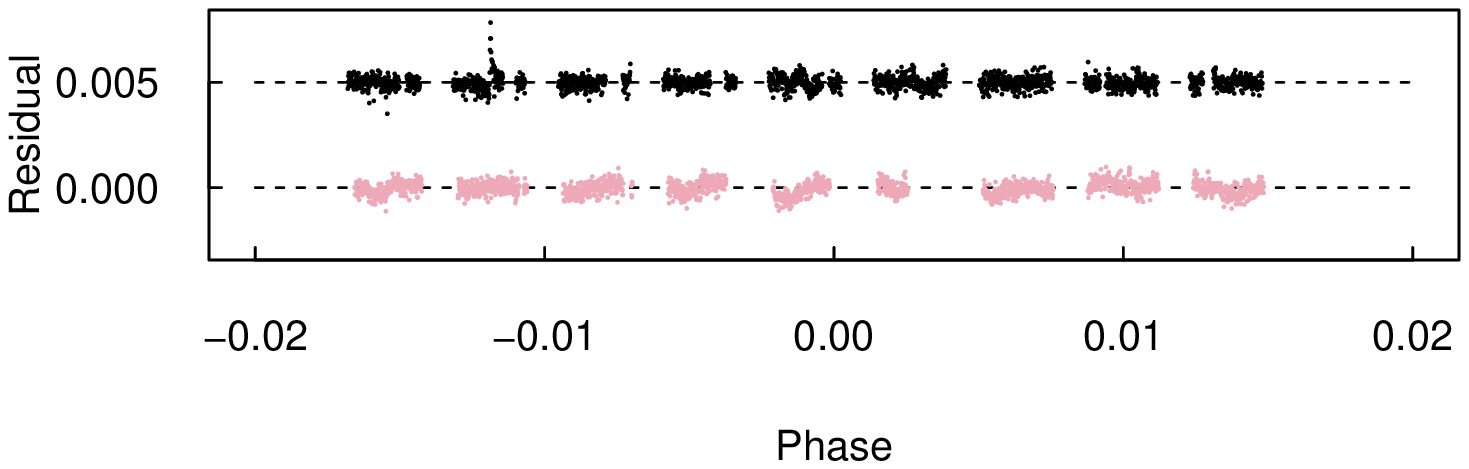}
    \caption{Best-fit transit solutions of the four \textit{CHEOPS} visits of AU\,Mic\,b (upper panels) and AU\,Mic\,c (lower panels) analyzed in this Letter, after omitting the flares during the transits. The phased points of the individual light curves are shown in light blue, and the binned light curve points of the individual transits are shown in dark blue. The lower panel shows the residuals observed at each individual visit. We highlight the increased variability of the light curve shape before the mid-transit of AU\,Mic\,b.}
    \label{fig:solutions}
\end{figure}

After masking out the flares (as shown in Fig. \ref{fig:rawLightCurves}), we determined the transit parameters using the \texttt{pycheops} software module \citep{2021MNRAS.tmp.3057M}. \texttt{pycheops} uses the \texttt{qpower2} transit model and the power-2 limb-darkening law \citep[][]{Maxted2019}; it calculates transit models of a spot-free star and a planet with a circular silhouette. {The model parameters are the transit time ($T_t$), the transit depth parameter $D = (R_p/R_\star)^2$, the transit duration parameter
$W=(R_\star/a)\sqrt{(1+k)^2 - b^2}/\pi$, and the impact parameter
$b = a \cos(i)/R_\star$. { The other system parameters can be derived from the above set; for example the transit duration expressed in hours as $W[h]=W\times P_i[h]$}, where $P_i$ is the instantaneous orbital period.} 

It is possible to fit more complex models to the observed transits that account for the presence of spots.
In order to constrain the spot modeling, we performed ground-based observations simultaneously with the \textit{CHEOPS} observations. We intend to publish this more complex modeling in a forthcoming paper, as the scope of this Letter is to primarily report the unexpectedly large TTVs, for which the standard modeling methods are sufficient.

%It uses a Gaussian Process (GP) regression to account for the roll angle systematics, whose parameters are estimated using the out-of-transit data. 

The priors we used are listed in Table~\ref{table:priors} for both AU\,Mic\,b and AU\,Mic\,c. {The stellar fundamental parameters were taken from SWEET-Cat \citep{2018A&A...620A..58S}, which are the same parameters as in \cite{2020Natur.582..497P}.} 
The noise model was calculated with \texttt{celerite} using the white-noise term \texttt{JitterTerm}$(\log \sigma_w)$ plus, optionally, a GP with kernel \texttt{SHOTerm}$(\log \omega _{0}, \log S_0, \log Q)$. The priors were identical to what we set up in \cite{2021A&A...654A.159S}.

The long-period trends of TTVs lead to the apparent change of the instantaneous orbital period. We fitted the instantaneous period as the $P_i$ parameter to remove the linear trend of TTV in 2021. The actual mean orbital period, $P_{\rm mean}$, is the one that minimizes the scatter of the TTV. $P_{\rm mean}$ was determined by an $O-C$ analysis of mid-transit times (see Sect. \ref{results}, and also Tables \ref{table:ttvsb} and \ref{table:ttvsc}).

\section{Results}
\label{results}

\begin{figure}
\includegraphics[viewport=0 207 490 480, width=\columnwidth]{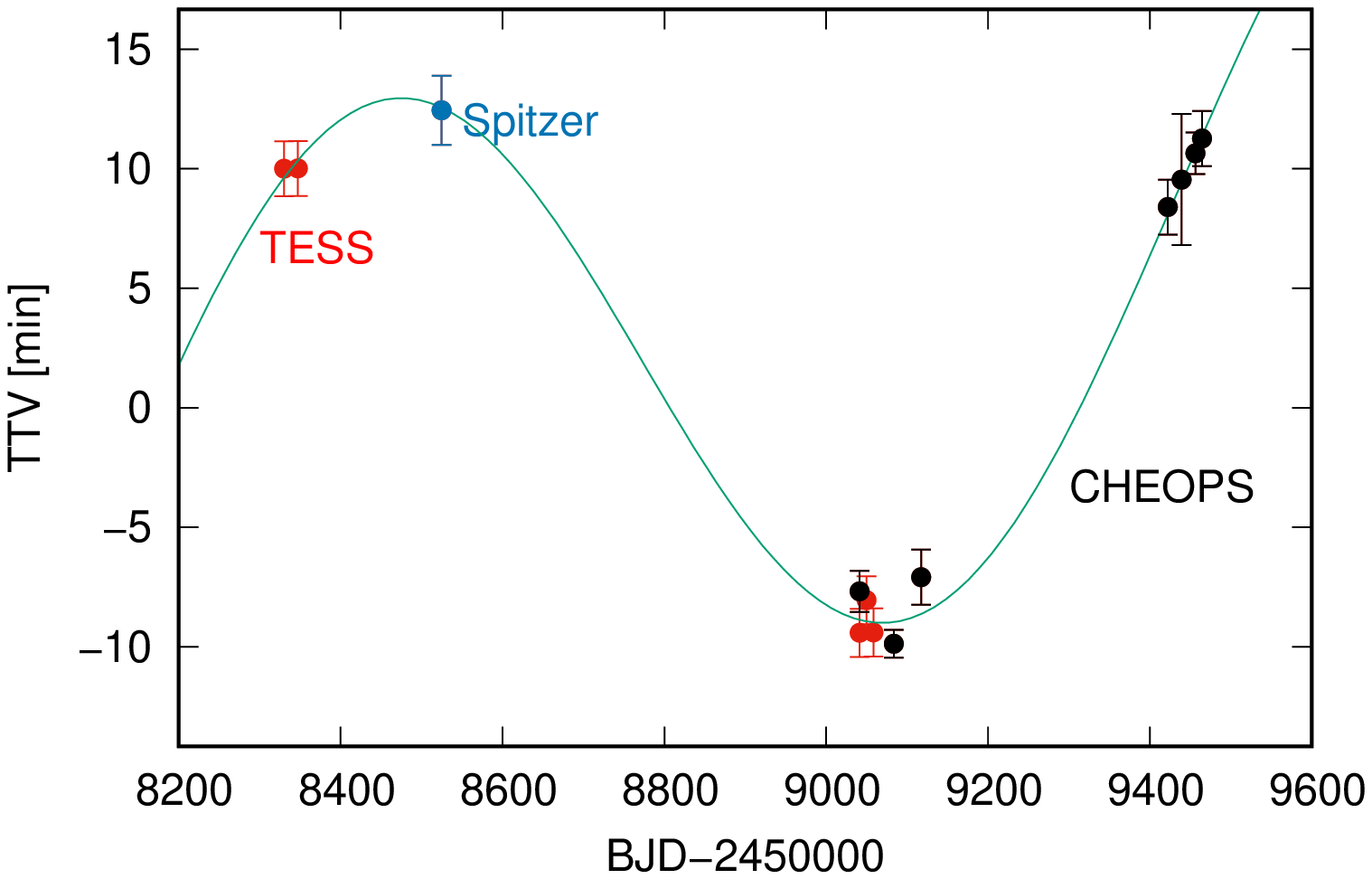}
\includegraphics[viewport=0 177 490 490, width=\columnwidth]{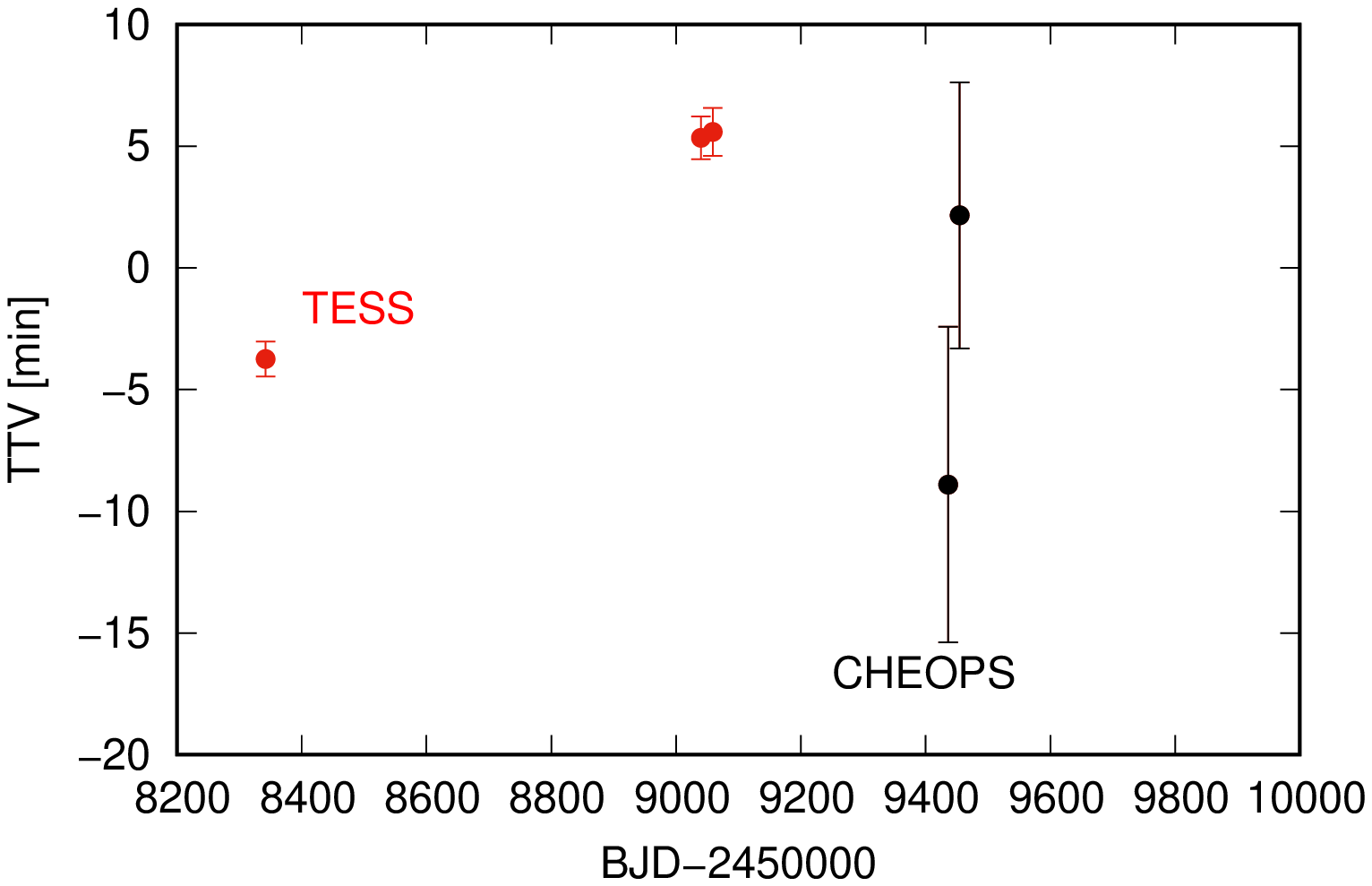}
\caption{TTV diagrams of AU\,Mic b (upper panel) calculated with $T_c=2\,458\,330.38416$ and $P_{\rm mean}=8.4631427$~d, and AU\,Mic c (lower panel) calculated with $T_c=9454.8973$ and $P_{\rm mean}=18.85882$~d. We included \textit{TESS} (red symbols), Spitzer (blue symbols), and \textit{CHEOPS} (black symbols) measurements. The harmonic fit to AU\,Mic\,b data illustrates the most probable shape of a periodic TTV fitted to the data. This is shown to illustrate the trend of the distribution without any dynamical interpretation.}
\label{fig:ttvs}
\end{figure}

\begin{figure}
\includegraphics[viewport=0 180 490 480, width=\columnwidth]{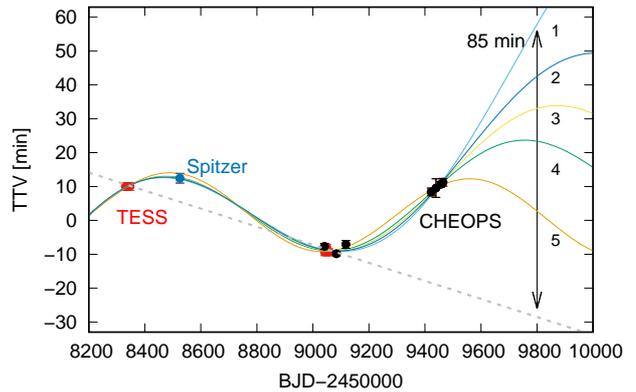}
\caption{Different predictions from the simplest harmonic TTV models to the 2022 opposition (colored curves) and the previously published linear ephemeris (gray dotted line). The colored curves fit equally well to all data points (see Table \ref{tab:coefficients} for their coefficients). Transits in 2022 August are expected to occur 40--85\,min later than predictions prior to the \textit{CHEOPS} 2021 observations.}
\label{fig:ttvs2}
\end{figure}

\begin{figure}
    \centering
\includegraphics[bb=0 65 440 385,width=\columnwidth]{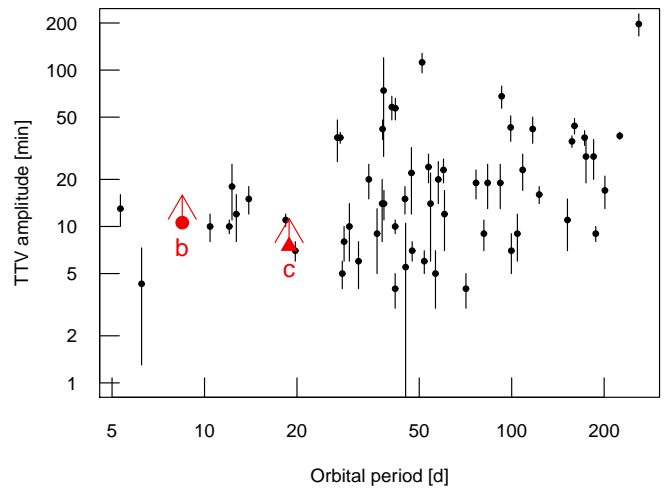}
    \caption{Half-amplitude of periodic TTVs observed for \textit{Kepler} planets (small dots) in comparison to the planets in the AU\,Mic system (red dot: AU\,Mic\,b, red triangle: AU\,Mic\,c). Due to the sparse coverage of observations, lower limits of TTV half-amplitudes are shown for AU\,Mic\,b and c.}
    \label{fig:ttvcomp}
\end{figure}

The best-fit transit parameters to 2021 \textit{CHEOPS} observations are summarizsed in Tables~\ref{tab:solutionsb} and \ref{tab:solutionsc}, both for AU\,Mic\,b and AU\,Mic\,c. We compare our results to previous estimates in \cite{2020Natur.582..497P}, \cite{2021A&A...654A.159S}, \cite{2021A&A...649A.177M}, and \cite{2021arXiv210903924G}. In \cite{2021A&A...654A.159S}, we derived three sets of solutions using different solving algorithms (\texttt{pycheops} \citealt{2021MNRAS.tmp.3057M} and \texttt{TLCM} \citealt{2020MNRAS.496.4442C}), with resulting solutions that did not differ significantly from each other. As the present Letter uses the \texttt{pycheops} algorithm, we selected the \texttt{pycheops} solution in \cite{2021A&A...654A.159S} for comparison with the 2021 observations.

All parameters with the exception of $R_p/R_\star$ (and hence $R_p$) are compatible with all previous solutions in the literature. The most significant improvement of the present analysis is the increased precision of the transit duration, $W$, thanks to the combined precision of the four \textit{CHEOPS} light curves, which have very good internal photometric accuracy.

The most relevant difference is observed in the $R_p/R_\star$ (and consequently, the $R_p$) planet radius parameters. In 2021, we observed smaller values than in 2018 and 2020, with a difference between the current and previous $R_p/R_\star$ estimates of 3.4$\sigma$. This is probably because the spot coverage of the stellar surface evolved between 2018 and 2021. It is known that spots along the transit chord decrease the transit depth, while unocculted spots increase it, and  the naively determined $R_p/R_\star$ size parameter becomes inconsistent  \citep{2013A&A...556A..19O}. Assuming that most spots are out of the transit chord, and therefore that spots {increase} the transit depth, \cite{2021A&A...649A.177M} derived that in 2020, the actual diameter was overestimated by about 6\%{} based on \textit{TESS} data. Our result shows that the activity level {increases} between 2020 and 2021 while the transit depth {decreases} significantly, rather than being biased to larger values. The most simple explanation would be that spots of the host star are mostly \textit{on the transit chord} of AU\,Mic\,b. We will discuss this possibility in detail in a forthcoming paper.

\begin{table*}
    \centering
    \caption{Best-fitting parameters of AU\,Mic\,b. {The parameters are compared to the results of \citet[P2020]{2020Natur.582..497P}, \citet[M2021]{2021A&A...649A.177M}, \citet[Sz2021]{2021A&A...654A.159S}, and \citet[G2021]{2021arXiv210903924G}.}}
    \label{tab:solutionsb}
    \begin{tabular}{llllll}
    \hline
    \noalign{\smallskip}
      & 
  This Letter &   
  P2020  &
  M2021  &
  Sz2021 &  
  G2021
  \\
      \noalign{\smallskip}
\hline
      \noalign{\smallskip}
      \hline
      \noalign{\smallskip}
$R_p/R_\star$  &
  0.0433$\pm$ 0.0017&
  0.0514$\pm$0.0013 & 
  0.0526$^{+0.0003}_{-0.0002}$ &
  0.0531 $\pm$ 0.0023  &     
  0.0512 $\pm$  0.0020   
  \\
$a/R_\star$ &  
  18.95 $\pm$ 0.35 &
  19.1$_{-1.6}^{+1.8}$ &
  19.1$^{+0.2}_{-0.4}$ &
  19.24 $\pm $   0.37 & 
  19.07 
 \\
$W$ [h]  &   
  3.51$\pm$ 0.03 &
  3.50$_{-0.59}^{+0.63}$ &
  3.50 $\pm$ 0.08 &
  3.48 $\pm $ 0.19 & 
  3.56 $_{-0.46}^{+0.60}$  \\
$R_p$ [R$_{\oplus}$] &
  3.55 0.13 &
  4.29$\pm $0.20 &
  4.07 $\pm$ 0.17 &
  4.36 $\pm $ 0.18 & 
  4.27 $\pm $ 0.17 
  \\
$a$ [AU] &
  0.0654 $\pm$0.0012 &
  0.066$_{-0.006}^{+0.007}$ &
  0.0645 $\pm$ 0.0013 &
  0.0678 $\pm $ 0.0013 & 
  0.0644 $_{-0.0054}^{+0.0056}$  \\
$b$  &  
  0.17 $\pm$ 0.11 &
  0.16$_{-0.11}^{+0.14}$ &
  0.18 $\pm$ 0.11&
  0.09 $\pm $  0.05 & 
  0.26 $_{-0.17}^{+0.13}$  \\ %
\noalign{\smallskip}
\hline
    \end{tabular}
\end{table*}

\begin{table*}
    \centering
    \caption{Best-fitting parameters of AU\,Mic\,c. {The parameters are compared to the results of M2021 and G2021.} }
    \label{tab:solutionsc}
    \begin{tabular}{llll}
    \hline
    \noalign{\smallskip}
      & 
  This Letter &   
  M2021  &
  G2021
  \\
      \noalign{\smallskip}
\hline
      \noalign{\smallskip}
      \hline
      \noalign{\smallskip}
$R_{\rm p}/R_\star$  &
  0.0313$\pm$0.0016&
  0.0395$\pm$0.0011 &   %
  0.0340$^{+0.0033}_{-0.0034}$ \\
$a/R_\star$ &  
  28.8 $\pm$ 2.4 &
  29$\pm$ 3.0 & %
  31.7 $^{+2.7}_{2.6}$
 \\
$W$ [h]  &   
  4.29$\pm$ 0.30 &
  4.50$\pm$ 0.80 & %
  4.20 $_{-0.67}^{+0.92}$  \\
$R_{\rm p}$ [R$_{\oplus}$] &
  2.56 $\pm$ 0.12 &
  3.24 $\pm$ 0.16 &  %
  2.79 $_{-0.30}^{+0.31}$  
  \\
$a$ [AU] &
  0.0993 $\pm$0.0085 &
  0.1101 $\pm$ 0.0022 & %
  0.110 $_{-0.010}^{+0.010}$  \\
$b$  &  
  0.58 $\pm$ 0.13 &
  0.51 $\pm$ 0.21& %
  0.30 $_{-0.20}^{+0.21}$  \\ %
\noalign{\smallskip}
\hline
    \end{tabular}
\end{table*}

\subsection{Large-amplitude TTVs}

Both planets show a very prominent TTV. We complemented the \textit{CHEOPS} (2020 and 2021) observations with mid-transit times of \textit{TESS} data (2018 and 2020), and also the mid-transit time of the single Spitzer measurement fitted by \cite{2020Natur.582..497P} (see the mid-transit times in Tables~\ref{table:ttvsb} and \ref{table:ttvsc}).
Following \cite{2012ApJ...761..122L}, the simplest form of a TTV can be described as a sinusoidal function with amplitude $A$ and superperiod $p$, which are complicated functions of the exoplanet system parameters. Therefore,
\begin{equation}
O-C = A \sin \bigg( 2 \pi {T_t \over p} + \phi \bigg) + c_1 T_t + c_2,
\label{oc:eq}
\end{equation}
where $T_t$ are individual transit times, and $\phi$, $c_1$, and $c_2$ are free parameters determined by the selection of the time coordinate and the trial orbital period. We derive the mean orbital period $P_{\rm mean}$ and $T_c$ , while demanding the elimination of the constant and the linear terms in $O-C$: $c_1=c_2=0$. (If there is no suspicion of a periodic TTV, linear epheremides (period and transit time) are given as A=0 solutions of Eq. \ref{oc:eq}.)
The currently available times of minima lead to $T_c = 2\,458\,330.38416\pm0.00005$~d and $P_{\rm mean} = 8.4631427 \pm 0.0000005$~d in case of AU\,Mic\,b and $T_c = 2\,4594\,54.8973\pm0.0005$~d and $P_{\rm mean} = 18.85882\pm0.00005$~d in case of  AU\,Mic\,c. 
With this selection, the peak-to-peak amplitude of the TTV of  AU\,Mic\,b and c are 23 and 9.5\,min, respectively (Fig. \ref{fig:ttvs}). 

The derived $P_{\rm mean}$ depends on which parts of the $O-C$ are constrained by data. It is likely that $P_{\rm mean}\approx P_{\rm orb}$, but, in general, equality cannot be guaranteed, in particular when the $O-C$ is sparsely sampled and with significant TTVs, as in the present case.
The result is that the fitted coefficients $c_1$ and $c_2$ become uncertain. Despite the fact that $T_c$ and $P_{\rm mean}$ were determined such that $c_1$ and $c_2$ vanish, they have error terms that correlate to the other parameters. The simplest form of an $O-C$ model describing a periodic process has all five free variables in the form of Eq.~\ref{oc:eq}: $A$, $P,$ and $\phi$ that are fitted, and $c_1$ and $c_2$ as parametrric constants determined from the data, which are therefore in correlation with the other three parameters.

The currently measured transits are concentrated in four relatively narrow windows (from a single transit to three-month observations) in the case of AU\,Mic\,b, and three similar windows in the case of AU\,Mic\,c. This leads to degeneracies in the best-fitting models of AU\,Mic\,b. Because of these degeneracies, we did not attempt to fit any curve to the current AU\,Mic\,c data. The future transit times extrapolated from the modeled $O-C$ curves of AU\,Mic\,b are wildly diverging. The $O-C$ (determined as described above) and the possible $O-C$ predictions for the future (being equivalently well-fitting solutions of Eq. \ref{oc:eq}) are plotted in Fig.~\ref{fig:ttvs2} with curves, and the gray dashed line shows the predictions from the linear ephemeris based on previously published data. Transits predicted for 2022 are expected to occur 40--85\,min later than expected without the 2021 observations of \textit{CHEOPS}. We note that if the dominant process in the long-term behavior of AU\,Mic\,b is a period change with a constant rate, and the appropriate fit is a parabola instead of a harmonic function, the difference in transit time can be even more than 90\,min. However, we think that this scenario is unlikely, because the position of the single measurement with the Spitzer telescope and the seemingly anticorrelated variations of the two planets in the system strongly suggest a periodic TTV in its leading term. 

This TTV {may} reflect orbital changes. \cite{2016A&A...585A..72I} suggested that starspots can also cause apparent TTVs of up to $\approx 1$\%{} of the transit duration. This would be on the order of 2\,min in the case of AU\,Mic\,b. {The observed effect is an order of magnitude larger than this prediction, and moreover with a pattern that is incompatible with the random behavior expected from stochastic spot occultations. Together, these findings  strongly suggest an orbital dynamics origin of the TTVs.}

The large TTVs of the AU\,Mic system are unusual, as can be seen when comparing to \textit{Kepler} planets with confirmed TTVs (Fig.~\ref{fig:ttvcomp}). Among the \textit{Kepler} planets, only two that exhibit TTVs have shorter orbital periods than AU\,Mic\,b \citep[Kepler-25\,b and Kepler-1530\,c, see][]{2019RAA....19...41G}. AU\,Mic\,b has a TTV semi-amplitude of $A \geq 11.5$\,min, which is large in comparison to other planets with known TTVs. AU\,Mic\,c is still at the short-period end of the planets with known TTVs, while the semi-amplitude of the TTV is known with lower precision.

{\cite{2021A&A...649A.177M} estimate that the AU\,Mic\,b--c interactions lead to a superperiod of $\approx$82 days and significantly smaller amplitude than reported here. While we are certain that the TTV has its origin in the orbital physics, there are still too few data points to conclusively determine a superperiod or the possibility of a linear period drift in addition to the periodic TTV. These effects could point to either a currently ongoing migration or additional perturbing outer planets. To address these questions, a longer time-span of observed transit timings is required.}

\section{Summary}

In this paper, we report our analysis of new \textit{CHEOPS} observations of both planets in the AU\,Mic system and draw the following main conclusions:

\begin{enumerate}

\item{} AU\,Mic\,b shows very significant TTVs, with a minimum-to-maximum amplitude $\geq23$\,min. AU\,Mic\,c shows TTVs with a minimum-to-maximum amplitude of $\geq 9.5$\,min. The best fitting mean orbital periods of AU\,Mic\,b and c are $P_{\rm mean}=8.4631427\pm0.0000005$~d and $P_{\rm mean}=18.85882\pm0.000005$~d, respectively. Taking the TTV into account, we predict that the transit times of AU\,Mic\,b in 2022 will happen 40--90\,min later than expected from previously published linear ephemeris.

\item{} The transit depths of both planets are observed by \textit{CHEOPS} to be smaller in 2021 than in 2020. The most likely reason is the increased activity of AU\,Mic with significant changes in the spot structure on the stellar surface. 

\item{} Due to the large influence of spots on the size parameter $R_p/R_\star$, its value should only be considered  as a proxy for the actual sizes of the planets. A de-biased size determination requires detailed spot modeling with contemporary complementary observations, which we will address in a forthcoming paper.

\end{enumerate}

The large-amplitude TTVs imply that the observations during the 2022 visibility have to be planned circumspectly. The ambiguity in transit-time predictions can be inaccurate up to 40--85\,min which is about half of the transit duration. This can be especially critical for scheduling follow-up observations of either planet, for example with the \textit{Hubble Space Telescope} or the \textit{James Webb Space Telescope}.

\section*{Acknowledgements}

\textit{CHEOPS} is an ESA mission in partnership with Switzerland with important contributions to the payload and the ground segment from Austria, Belgium, France, Germany, Hungary, Italy, Portugal, Spain, Sweden, and the United Kingdom. The CHEOPS Consortium would like to gratefully acknowledge the support received by all the agencies, offices, universities, and industries involved. Their flexibility and willingness to explore new approaches were essential to the success of this mission. 
GyMSz and ZG acknowledge the support of the Hungarian National Research, Development and Innovation Office (NKFIH) grant K-125015, a a PRODEX Experiment Agreement No. 4000137122 between the ELTE E\"otv\"os Lor\'and University and the European Space Agency (ESA-D/SCI-LE-2021-0025), the Lend\"ulet LP2018-7/2021 grant of the Hungarian Academy of Science and the support of the city of Szombathely. ZG was supported by the VEGA grant of the Slovak Academy of Sciences
No. 2/0031/22 and by the Slovak Research and Development Agency - the
contract No. APVV-20-0148.
ABr was supported by the SNSA. 
DG gratefully acknowledges financial support from the CRT foundation under Grant No. 2018.2323 ``Gaseous or rocky? Unveiling the nature of small worlds''. 
ACC and TW acknowledge support from STFC consolidated grant number ST/M001296/1. 
A.De. acknowledges support from the European Research Council (ERC) under the European Union's Horizon 2020 research and innovation programme (project {\sc Four Aces}, grant agreement No. 724427), and from the National Centre for Competence in Research ``PlanetS'' supported by the Swiss National Science Foundation (SNSF).
This work was also partially supported by a grant from the Simons Foundation (PI Queloz, grant number 327127). 
ML acknowledges support of the Swiss National Science Foundation under grant number PCEFP2\_{}194576. 
GSc, GPi, IPa, LBo, RRa, and VNa and RRa acknowledge the funding support from Italian Space Agency (ASI) regulated by``Accordo ASI-INAF n. 2013-016-R.0 del 9 luglio 2013 e integrazione del 9 luglio 2015 CHEOPS Fasi A/B/C''. 
LMS gratefully acknowledges financial support from the CRT foundation under Grant No. 2018.2323 ``Gaseous or rocky? Unveiling the nature of small worlds''. 
V.V.G. is an F.R.S-FNRS Research Associate. 
YA and MJH acknowledge the support of the Swiss National Fund under grant 200020\_172746. 
We acknowledge support from the Spanish Ministry of Science and Innovation and the European Regional Development Fund through grants ESP2016-80435-C2-1-R, ESP2016-80435-C2-2-R, PGC2018-098153-B-C33, PGC2018-098153-B-C31, ESP2017-87676-C5-1-R, MDM-2017-0737 Unidad de Excelencia Maria de Maeztu-Centro de Astrobiologí­a (INTA-CSIC), as well as the support of the Generalitat de Catalunya/CERCA programme. The MOC activities have been supported by the ESA contract No. 4000124370. 
S.C.C.B. acknowledges support from FCT through FCT contracts nr. IF/01312/2014/CP1215/CT0004. 
XB, SC, DG, MF and JL acknowledge their role as ESA-appointed CHEOPS science team members. 
This project was supported by the CNES. 
The Belgian participation to CHEOPS has been supported by the Belgian Federal Science Policy Office (BELSPO) in the framework of the PRODEX Program, and by the University of Liège through an ARC grant for Concerted Research Actions financed by the Wallonia-Brussels Federation. 
This work was supported by FCT - Fundação para a Ciência e a Tecnologia through national funds and by FEDER through COMPETE2020 - Programa Operacional Competitividade e Internacionalizacão by these grants: UID/FIS/04434/2019, UIDB/04434/2020, UIDP/04434/2020, PTDC/FIS-AST/32113/2017 \& POCI-01-0145-FEDER- 032113, PTDC/FIS-AST/28953/2017 \& POCI-01-0145-FEDER-028953, PTDC/FIS-AST/28987/2017 \& POCI-01-0145-FEDER-028987, O.D.S.D. is supported in the form of work contract (DL 57/2016/CP1364/CT0004) funded by national funds through FCT. 
B.-O.D. acknowledges support from the Swiss National Science Foundation (PP00P2-190080). 
This project has received funding from the European Research Council (ERC) under the European Union’s Horizon 2020 research and innovation programme (project {\sc Four Aces} grant agreement No 724427).   It has also been carried out in the frame of the National Centre for Competence in Research PlanetS supported by the Swiss National Science Foundation (SNSF). DE acknowledges financial support from the Swiss National Science Foundation for project 200021\_{}200726.
LD is an F.R.S.-FNRS Postdoctoral Researcher.
MF and CMP gratefully acknowledge the support of the Swedish National Space Agency (DNR 65/19, 174/18). 
M.G. is an F.R.S.-FNRS Senior Research Associate. 
SH gratefully acknowledges CNES funding through the grant 837319. 
KGI is the ESA CHEOPS Project Scientist and is responsible for the ESA CHEOPS Guest Observers Programme. She does not participate in, or contribute to, the definition of the Guaranteed Time Programme of the CHEOPS mission through which observations described in this paper have been taken, nor to any aspect of target selection for the programme. 
This work was granted access to the HPC resources of MesoPSL financed by the Region Ile de France and the project Equip@Meso (reference ANR-10-EQPX-29-01) of the programme Investissements d'Avenir supervised by the Agence Nationale pour la Recherche. 
PM acknowledges support from STFC research grant number ST/M001040/1. 
Acknowledges support from the Spanish Ministry of Science and Innovation and the European Regional Development Fund through grant PGC2018-098153-B- C33, as well as the support of the Generalitat de Catalunya/CERCA programme. 
S.G.S. acknowledge support from FCT through FCT contract nr. CEECIND/00826/2018 and POPH/FSE (EC).

\bibliographystyle{aa}
\bibliography{aumic}

\begin{thebibliography}{20}
\expandafter\ifx\csname natexlab\endcsname\relax\def\natexlab#1{#1}\fi

\bibitem[{{Benz} {et~al.}(2021){Benz}, {Broeg}, {Fortier}, {Rando}, {Beck},
  {Beck}, {Queloz}, {Ehrenreich}, {Maxted}, {Isaak}, {Billot}, {Alibert},
  {Alonso}, {Ant{\'o}nio}, {Asquier}, {Bandy}, {B{\'a}rczy}, {Barrado},
  {Barros}, {Baumjohann}, {Bekkelien}, {Bergomi}, {Biondi}, {Bonfils},
  {Borsato}, {Brandeker}, {Busch}, {Cabrera}, {Cessa}, {Charnoz}, {Chazelas},
  {Collier Cameron}, {Corral Van Damme}, {Cortes}, {Davies}, {Deleuil},
  {Deline}, {Delrez}, {Demangeon}, {Demory}, {Erikson}, {Farinato}, {Fossati},
  {Fridlund}, {Futyan}, {Gandolfi}, {Garcia Munoz}, {Gillon}, {Guterman},
  {Gutierrez}, {Hasiba}, {Heng}, {Hernandez}, {Hoyer}, {Kiss}, {Kovacs},
  {Kuntzer}, {Laskar}, {Lecavelier des Etangs}, {Lendl}, {L{\'o}pez}, {Lora},
  {Lovis}, {L{\"u}ftinger}, {Magrin}, {Malvasio}, {Marafatto}, {Michaelis}, {de
  Miguel}, {Modrego}, {Munari}, {Nascimbeni}, {Olofsson}, {Ottacher},
  {Ottensamer}, {Pagano}, {Palacios}, {Pall{\'e}}, {Peter}, {Piazza}, {Piotto},
  {Pizarro}, {Pollaco}, {Ragazzoni}, {Ratti}, {Rauer}, {Ribas}, {Rieder},
  {Rohlfs}, {Safa}, {Salatti}, {Santos}, {Scandariato}, {S{\'e}gransan},
  {Simon}, {Smith}, {Sordet}, {Sousa}, {Steller}, {Szab{\'o}}, {Szoke},
  {Thomas}, {Tschentscher}, {Udry}, {Van Grootel}, {Viotto}, {Walter},
  {Walton}, {Wildi}, \& {Wolter}}]{2021ExA....51..109B}
{Benz}, W., {Broeg}, C., {Fortier}, A., {et~al.} 2021, Experimental Astronomy,
  51, 109

\bibitem[{{Csizmadia}(2020)}]{2020MNRAS.496.4442C}
{Csizmadia}, S. 2020, \mnras, 496, 4442

\bibitem[{{Gaia Collaboration} {et~al.}(2018){Gaia Collaboration}, {Brown},
  {Vallenari}, {Prusti}, {de Bruijne}, {Babusiaux}, {Bailer-Jones}, {Biermann},
  {Evans}, {Eyer}, \& et~al.}]{2018A&A...616A...1G}
{Gaia Collaboration}, {Brown}, A.~G.~A., {Vallenari}, A., {et~al.} 2018, \aap,
  616, A1

\bibitem[{{Gajdo{\v{s}}} {et~al.}(2019){Gajdo{\v{s}}}, {Va{\v{n}}ko}, \&
  {Parimucha}}]{2019RAA....19...41G}
{Gajdo{\v{s}}}, P., {Va{\v{n}}ko}, M., \& {Parimucha}, {\v{S}}. 2019, Research
  in Astronomy and Astrophysics, 19, 041

\bibitem[{{Gilbert} {et~al.}(2021){Gilbert}, {Barclay}, {Quintana},
  {Walkowicz}, {Vega}, {Schlieder}, {Monsue}, {Cale}, {Collins}, {Gaidos}, {El
  Mufti}, {Reefe}, {Plavchan}, {Tanner}, {Wittenmyer}, {Wittrock}, {Jenkins},
  {Latham}, {Ricker}, {Rose}, {Seager}, {Vanderspek}, \&
  {Winn}}]{2021arXiv210903924G}
{Gilbert}, E.~A., {Barclay}, T., {Quintana}, E.~V., {et~al.} 2021, arXiv
  e-prints, arXiv:2109.03924

\bibitem[{{Hoyer} {et~al.}(2020){Hoyer}, {Guterman}, {Demangeon}, {Sousa},
  {Deleuil}, {Meunier}, \& {Benz}}]{2020A&A...635A..24H}
{Hoyer}, S., {Guterman}, P., {Demangeon}, O., {et~al.} 2020, \aap, 635, A24

\bibitem[{{Ioannidis} {et~al.}(2016){Ioannidis}, {Huber}, \&
  {Schmitt}}]{2016A&A...585A..72I}
{Ioannidis}, P., {Huber}, K.~F., \& {Schmitt}, J.~H.~M.~M. 2016, \aap, 585, A72

\bibitem[{{Kiraga}(2012)}]{2012AcA....62...67K}
{Kiraga}, M. 2012, \actaa, 62, 67

\bibitem[{{Lithwick} {et~al.}(2012){Lithwick}, {Xie}, \&
  {Wu}}]{2012ApJ...761..122L}
{Lithwick}, Y., {Xie}, J., \& {Wu}, Y. 2012, \apj, 761, 122

\bibitem[{{Mamajek} \& {Bell}(2014)}]{2014MNRAS.445.2169M}
{Mamajek}, E.~E. \& {Bell}, C. P.~M. 2014, \mnras, 445, 2169

\bibitem[{{Martioli} {et~al.}(2021){Martioli}, {H{\'e}brard}, {Correia},
  {Laskar}, \& {Lecavelier des Etangs}}]{2021A&A...649A.177M}
{Martioli}, E., {H{\'e}brard}, G., {Correia}, A.~C.~M., {Laskar}, J., \&
  {Lecavelier des Etangs}, A. 2021, \aap, 649, A177

\bibitem[{{Maxted} {et~al.}(2021){Maxted}, {Ehrenreich}, {Wilson}, {Alibert},
  {Collier Cameron}, {Hoyer}, {Sousa}, {Olofsson}, {Bekkelien}, {Deline},
  {Delrez}, {Bonfanti}, {Borsato}, {Alonso}, {Anglada Escud{\'e}}, {Barrado},
  {Barros}, {Baumjohann}, {Beck}, {Beck}, {Benz}, {Billot}, {Biondi},
  {Bonfils}, {Brandeker}, {Broeg}, {B{\'a}rczy}, {Cabrera}, {Charnoz}, {Corral
  Van Damme}, {Csizmadia}, {Davies}, {Deleuil}, {Demangeon}, {Demory},
  {Erikson}, {Flor{\'e}n}, {Fortier}, {Fossati}, {Fridlund}, {Futyan},
  {Gandolfi}, {Gillon}, {Guedel}, {Guterman}, {Heng}, {Isaak}, {Kiss},
  {Laskar}, {Lecavelier des Etangs}, {Lendl}, {Lovis}, {Magrin}, {Nascimbeni},
  {Ottensamer}, {Pagano}, {Pall{\'e}}, {Peter}, {Piotto}, {Pollacco},
  {Pozuelos}, {Queloz}, {Ragazzoni}, {Rando}, {Rauer}, {Reimers}, {Ribas},
  {Santos}, {Scandariato}, {Simon}, {Smith}, {Steller}, {Swayne}, {Szab{\'o}},
  {S{\'e}gransan}, {Thomas}, {Udry}, {Van Grootel}, \&
  {Walton}}]{2021MNRAS.tmp.3057M}
{Maxted}, P.~F.~L., {Ehrenreich}, D., {Wilson}, T.~G., {et~al.} 2021, \mnras
  [\eprint[arXiv]{2111.08828}]

\bibitem[{{Maxted} \& {Gill}(2019)}]{Maxted2019}
{Maxted}, P.~F.~L. \& {Gill}, S. 2019, \aap, 622, A33

\bibitem[{{Miret-Roig} {et~al.}(2020){Miret-Roig}, {Galli}, {Brandner}, {Bouy},
  {Barrado}, {Olivares}, {Antoja}, {Romero-G{\'o}mez}, {Figueras}, \&
  {Lillo-Box}}]{2020A&A...642A.179M}
{Miret-Roig}, N., {Galli}, P.~A.~B., {Brandner}, W., {et~al.} 2020, \aap, 642,
  A179

\bibitem[{{Oshagh} {et~al.}(2013){Oshagh}, {Santos}, {Boisse}, {Bou{\'e}},
  {Montalto}, {Dumusque}, \& {Haghighipour}}]{2013A&A...556A..19O}
{Oshagh}, M., {Santos}, N.~C., {Boisse}, I., {et~al.} 2013, \aap, 556, A19

\bibitem[{{Plavchan} {et~al.}(2020){Plavchan}, {Barclay}, {Gagn{\'e}}, {Gao},
  {Cale}, {Matzko}, {Dragomir}, {Quinn}, {Feliz}, {Stassun}, {Crossfield},
  {Berardo}, {Latham}, {Tieu}, {Anglada-Escud{\'e}}, {Ricker}, {Vanderspek},
  {Seager}, {Winn}, {Jenkins}, {Rinehart}, {Krishnamurthy}, {Dynes}, {Doty},
  {Adams}, {Afanasev}, {Beichman}, {Bottom}, {Bowler}, {Brinkworth}, {Brown},
  {Cancino}, {Ciardi}, {Clampin}, {Clark}, {Collins}, {Davison},
  {Foreman-Mackey}, {Furlan}, {Gaidos}, {Geneser}, {Giddens}, {Gilbert},
  {Hall}, {Hellier}, {Henry}, {Horner}, {Howard}, {Huang}, {Huber}, {Kane},
  {Kenworthy}, {Kielkopf}, {Kipping}, {Klenke}, {Kruse}, {Latouf}, {Lowrance},
  {Mennesson}, {Mengel}, {Mills}, {Morton}, {Narita}, {Newton}, {Nishimoto},
  {Okumura}, {Palle}, {Pepper}, {Quintana}, {Roberge}, {Roccatagliata},
  {Schlieder}, {Tanner}, {Teske}, {Tinney}, {Vanderburg}, {von Braun}, {Walp},
  {Wang}, {Wang}, {Weigand }, {White}, {Wittenmyer}, {Wright}, {Youngblood},
  {Zhang}, \& {Zilberman}}]{2020Natur.582..497P}
{Plavchan}, P., {Barclay}, T., {Gagn{\'e}}, J., {et~al.} 2020, \nat, 582, 497

\bibitem[{{Ricker} {et~al.}(2014){Ricker}, {Winn}, {Vanderspek}, {Latham},
  {Bakos}, {Bean}, {Berta-Thompson}, {Brown}, {Buchhave}, {Butler}, {Butler},
  {Chaplin}, {Charbonneau}, {Christensen-Dalsgaard}, {Clampin}, {Deming},
  {Doty}, {De Lee}, {Dressing}, {Dunham}, {Endl}, {Fressin}, {Ge}, {Henning},
  {Holman}, {Howard}, {Ida}, {Jenkins}, {Jernigan}, {Johnson}, {Kaltenegger},
  {Kawai}, {Kjeldsen}, {Laughlin}, {Levine}, {Lin}, {Lissauer}, {MacQueen},
  {Marcy}, {McCullough}, {Morton}, {Narita}, {Paegert}, {Palle}, {Pepe},
  {Pepper}, {Quirrenbach}, {Rinehart}, {Sasselov}, {Sato}, {Seager},
  {Sozzetti}, {Stassun}, {Sullivan}, {Szentgyorgyi}, {Torres}, {Udry}, \&
  {Villasenor}}]{2014SPIE.9143E..20R}
{Ricker}, G.~R., {Winn}, J.~N., {Vanderspek}, R., {et~al.} 2014, in Society of
  Photo-Optical Instrumentation Engineers (SPIE) Conference Series, Vol. 9143,
  Space Telescopes and Instrumentation 2014: Optical, Infrared, and Millimeter
  Wave, ed. J.~{Oschmann}, Jacobus~M., M.~{Clampin}, G.~G. {Fazio}, \& H.~A.
  {MacEwen}, 914320

\bibitem[{{Sousa} {et~al.}(2018){Sousa}, {Adibekyan}, {Delgado-Mena}, {Santos},
  {Andreasen}, {Ferreira}, {Tsantaki}, {Barros}, {Demangeon}, {Israelian},
  {Faria}, {Figueira}, {Mortier}, {Brand{\~a}o}, {Montalto}, {Rojas-Ayala}, \&
  {Santerne}}]{2018A&A...620A..58S}
{Sousa}, S.~G., {Adibekyan}, V., {Delgado-Mena}, E., {et~al.} 2018, \aap, 620,
  A58

\bibitem[{{Szab{\'o}} {et~al.}(2021){Szab{\'o}}, {Gandolfi}, {Brandeker},
  {Csizmadia}, {Garai}, {Billot}, {Broeg}, {Ehrenreich}, {Fortier}, {Fossati},
  {Hoyer}, {Kiss}, {Lecavelier des Etangs}, {Maxted}, {Ribas}, {Alibert},
  {Alonso}, {Anglada Escud{\'e}}, {B{\'a}rczy}, {Barros}, {Barrado},
  {Baumjohann}, {Beck}, {Beck}, {Bekkelien}, {Bonfils}, {Benz}, {Borsato},
  {Busch}, {Cabrera}, {Charnoz}, {Collier Cameron}, {Van Damme}, {Davies},
  {Delrez}, {Deleuil}, {Demangeon}, {Demory}, {Erikson}, {Fridlund}, {Futyan},
  {Garc{\'\i}a Mu{\~n}oz}, {Gillon}, {Guedel}, {Guterman}, {Heng}, {Isaak},
  {Lacedelli}, {Laskar}, {Lendl}, {Lovis}, {Luntzer}, {Magrin}, {Nascimbeni},
  {Olofsson}, {Osborn}, {Ottensamer}, {Pagano}, {Pall{\'e}}, {Peter}, {Piazza},
  {Piotto}, {Pollacco}, {Queloz}, {Ragazzoni}, {Rando}, {Rauer}, {Santos},
  {Scandariato}, {S{\'e}gransan}, {Serrano}, {Sicilia}, {Simon}, {Smith},
  {Sousa}, {Steller}, {Thomas}, {Udry}, {Van Grootel}, {Walton}, \&
  {Wilson}}]{2021A&A...654A.159S}
{Szab{\'o}}, G.~M., {Gandolfi}, D., {Brandeker}, A., {et~al.} 2021, \aap, 654,
  A159

\bibitem[{{Torres} {et~al.}(2006){Torres}, {Quast}, {da Silva}, {de La Reza},
  {Melo}, \& {Sterzik}}]{2006A&A...460..695T}
{Torres}, C.~A.~O., {Quast}, G.~R., {da Silva}, L., {et~al.} 2006, \aap, 460,
  695

\end{thebibliography}

%%%APPENDIX starts from here

\appendix
\section{Data tables}

In this Appendix, we present tables with detailed parameters used in the transit fitting, as cited in the main text. 
%In Fig. \ref{riverplot}, a riverplot-like representation of the AU Mic b and c transits are shown.

%\begin{figure}
%    \centering
%    \includegraphics[bb=87 10 377 471, clip,width=0.5\columnwidth]{riverplot_b.eps}%
%    \includegraphics[bb=87 10 377 471, clip,width=0.5\columnwidth]{riverplot_c.eps}%
%    \caption{Riverplot-like representations of AU Mic b and c transits, built up by the transit models with parameters taken from \cite{2021arXiv210903924G,2021A&A...649A.177M} and this paper. The transit    models are phased according to the mean linear ephemeris and the curves are shifted to the BJD date of the observation.}
%    \label{fig:riverplot}
%\end{figure}

\begin{figure*}
    \centering
    \includegraphics[viewport=13 100 650 630,clip, width=16cm]{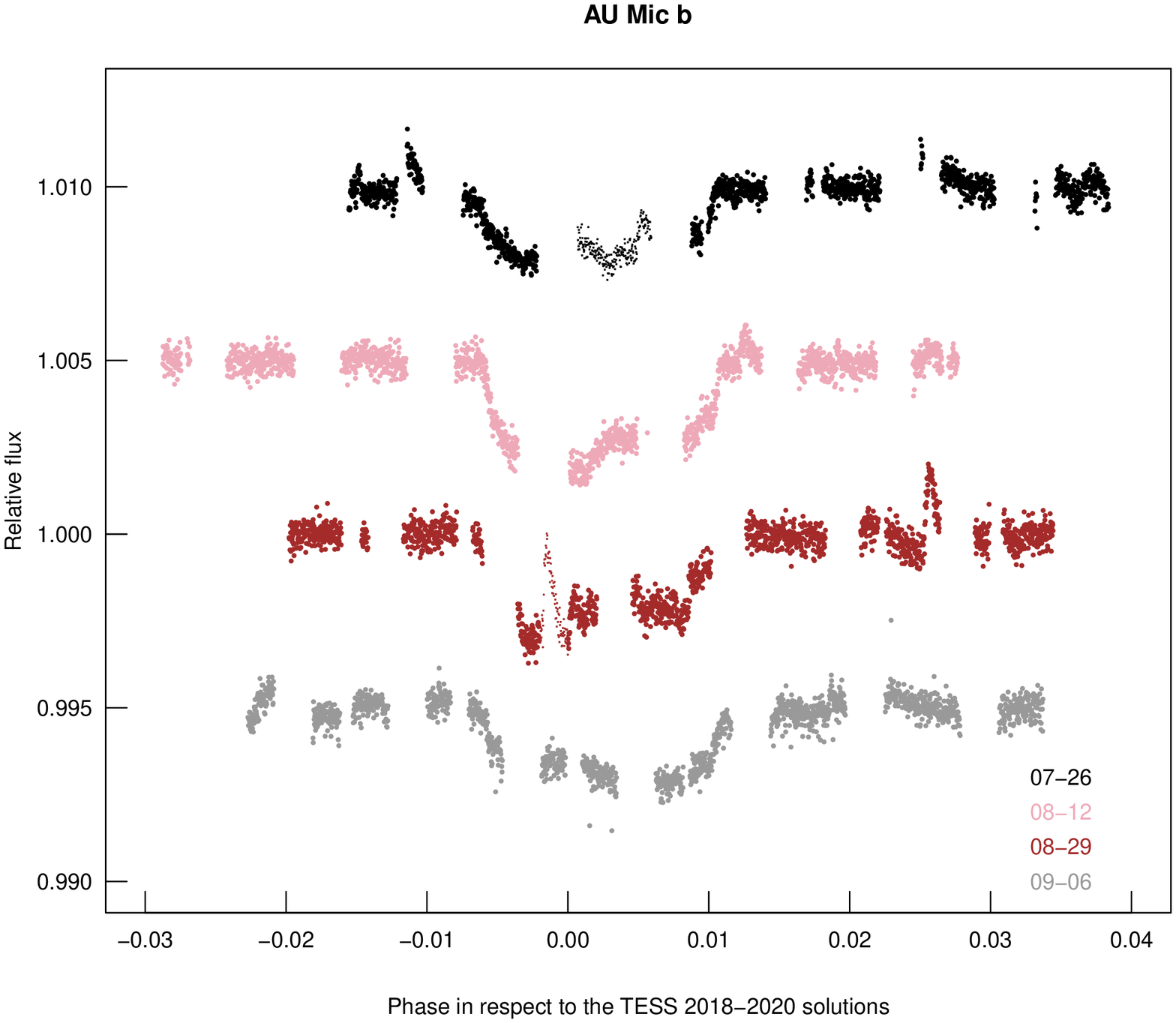}
    \includegraphics[viewport=13 200 650 500, width=16cm,clip]{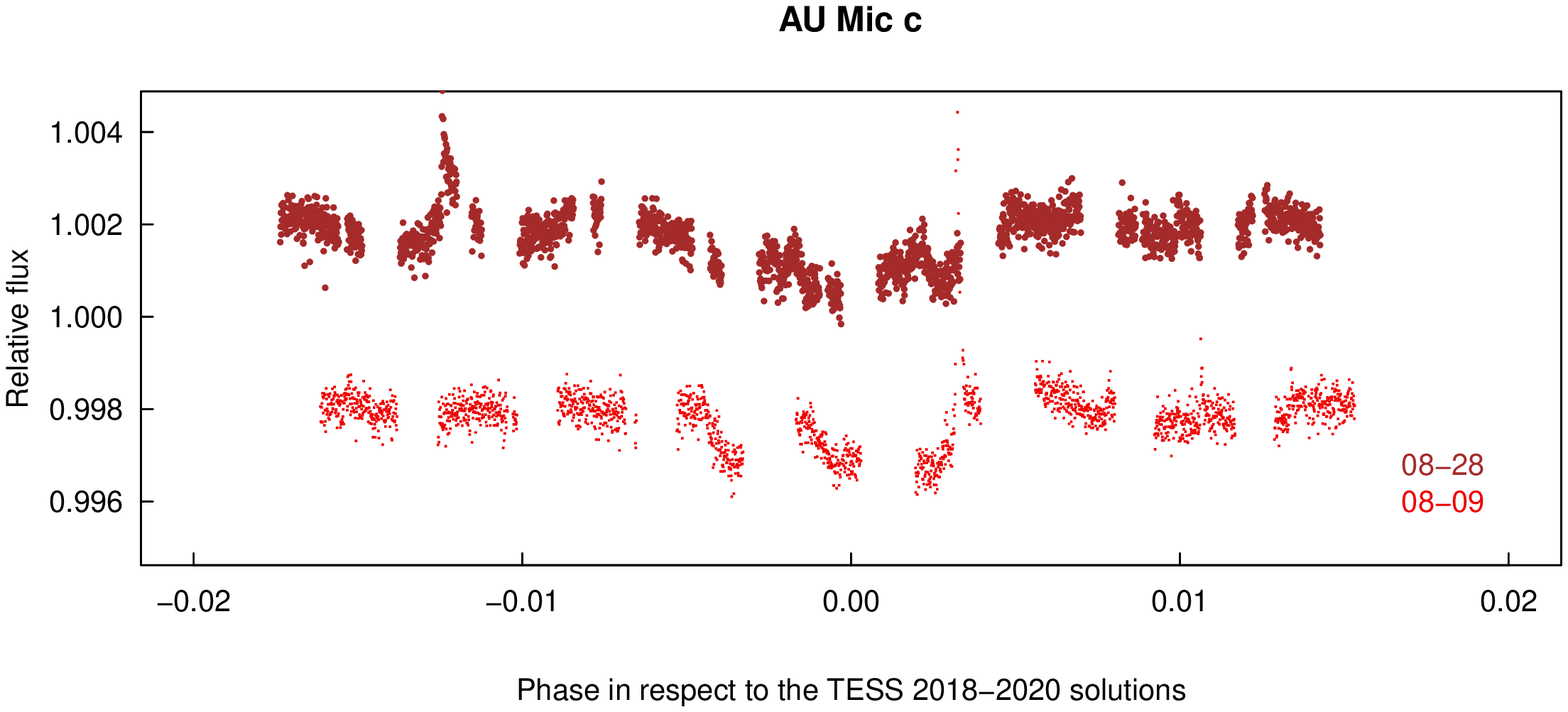}
    \caption{\textit{CHEOPS} observations of AU\,Mic\,b (upper panel) and AU\,Mic\,c (lower panel) transits analyzed in this paper. The  21-07-26 and 21-08-29 transits of AU\,Mic b were contaminated by flares. These points have been omitted from light curve fitting (plotted here with smaller dots). The labels show the date of the start of the visits in MM-DD format. The pixel/flux scale of the two panels is equal in order to show the relative amplitudes of the two planets.}
    \label{fig:rawLightCurves}
\end{figure*}

\begin{table}[]
    \centering
\caption{Priors and posteriors on planet parameters. We note that $P_i$ refers to the instantaneous period that best fits the 2021 measurements, and is different from $P_{\rm mean}$, which minimizes the scatter of the TTVs of all observations between 2018--2021.}
    \begin{tabular}{lll}
%\hline
     & \textbf{AU\,Mic\,b} & \textbf{AU\,Mic\,c} \\
\hline
\multicolumn{3}{c}{PRIORS~~~} \\
\hline
$D$&  $\mathcal{N}(0.003, 0.001)$ & $\mathcal{N}(0.0015, 0.0004)$ \\
$W[h]$&  $\mathcal{N}(3.656, 0.203)$ & $\mathcal{N}(3.892, 0.4)$ \\
$b$&   $\mathcal{U}(0.00, 0.30)$ & $\mathcal{U}(0.3, 0.7)$\\
$P_i[d]$& $\mathcal{N}(8.463000, 0.003)$ & 
       $\mathcal{N}(18.8590, 0.0030)$ \\
$T_0$& $\mathcal{N}(9447.52630, 0.004)$ & $\mathcal{N}(9436.0350, 0.0072)$\\
\hline
\multicolumn{3}{c}{POSTERIORS~~~} \\
\hline
$D$ & $0.00187\pm0.00008$& $0.00098\pm0.00010$\\
$W[h]$ & $3.514\pm0.026$ & $4.290\pm0.307$\\
$b$ & $0.17\pm0.10 $ & $0.58\pm0.13$\\
$P_i[d]$ & $8.46353\pm 0.00024$ & $18.8602\pm0.0026$ \\ 
$T_0$ & $9447.52634\pm 0.00046$& $9436.036\pm0.004$\\
%294ppm, 279 ppm
\hline
    \end{tabular}
    \label{table:priors}
\end{table}

\begin{table}[]
    \centering
\caption{Observed mid-transit times and $O-C$ values of AU\,Mic\,b based on \textit{TESS}, Spitzer, and
\textit{CHEOPS} observations analyzed in the present work, with $T_c = 2\,458\,330.38416$\,d and $P_{\rm mean} = 8.4631427$\,d.  References to transit times are: $a$: \cite{2021A&A...654A.159S}, 
$b$: \cite{2020Natur.582..497P},
$c$: This Letter. }
    \begin{tabular}{llrr}
\hline
Designation & Transit Time & $O-C$ & Err \\
            &  [BJD$-$2\,450\,000]  & [min] & [min] \\
\hline
\textit{TESS} S1\#{}1$^a$& 8330.3911$\pm$0.0009 & 10.00& 1.33 \\
\textit{TESS} S1\#{}2$^a$& 8347.3174$\pm$0.0009 & 10.02& 1.33\\
Spitzer\#{}1$^b$ & 8525.04509$\pm$0.0010 &12.45 & 1.43\\
\textit{TESS} S27\#{}1$^a$& 9041.2816$\pm$0.0008 &-9.42&1.17\\
\textit{TESS} S27\#{}2$^a$& 9049.7457$\pm$0.0008 &-8.05&1.17\\
\textit{TESS} S27\#{}3$^a$& 9058.2080$\pm$0.0008 &-9.40&1.17 \\
\textit{CHEOPS} 20-07-10$^a$& 9041.2828$\pm$0.0006 &-7.70 & 0.87 \\
\textit{CHEOPS} 20-08-21$^a$& 9083.5970$\pm$0.0004 &-9.88 & 0.58\\
\textit{CHEOPS} 20-09-24$^a$& 9117.4515$\pm$0.0008 &-7.08 & 1.17\\
\textit{CHEOPS} 21-07-26$^c$& 9422.1342$\pm$0.0010 & 8.40&1.43\\
\textit{CHEOPS} 21-08-12$^c$& 9439.0636$\pm$0.0021 & 9.55&3.15\\
\textit{CHEOPS} 21-08-29$^c$& 9455.9895$\pm$0.0007 & 10.65&1.0\\
\textit{CHEOPS} 21-09-06$^c$& 9464.4531$\pm$0.0009 & 11.25&1.27\\
\hline
    \end{tabular}
    \label{table:ttvsb}
\end{table}

\begin{table}[]
    \centering
\caption{Observed mid-transit times and $O-C$ values of AU\,Mic\,c based on \textit{TESS} and \textit{CHEOPS} observations analyzed in the present work, with $T_c = 2\,459\,454.8973$\,d and $P_{\rm mean} = 18.85882$\,d.  References to transit times are: $a$: \cite{2021arXiv210903924G}, $b$: This Letter. }
    \begin{tabular}{llrr}
\hline
Designation & Transit Time & O-C & Err \\
            &  [BJD$-$2\,450\,000]  & [min] & [min] \\
\hline
\textit{TESS} S1\#{}1$^a$& 8342.22432$\pm$0.0004 & -3.75 & 0.71 \\
\textit{TESS} S27\#{}1$^a$& 9040.00697$\pm$0.0005 & 5.34&0.88\\
\textit{TESS} S27\#{}2$^a$& 9058.86596$\pm$0.0006 & 5.58&0.98\\
\textit{CHEOPS} 21-08-09$^b$& 9436.0323$\pm$0.0045 & -8.90&6.48\\
\textit{CHEOPS} 21-08-28$^b$& 9454.8988$\pm$0.0038 & 2.16&5.47\\
\hline
    \end{tabular}
    \label{table:ttvsc}
\end{table}

\subsection{Coefficients of the model curves in Fig. \ref{fig:ttvs2}}

In this subsection, we give the coefficients of the model curves in Fig. \ref{fig:ttvs2}. The purpose of this table is to enable the reproduction of the figure. The coefficients refer to time measured in BJD-2\,450\,000 days. We emphasize that we do not attribute any physical interpretation to these fits. Moreover, the various fits are equally consistent with the current data.

The equation of the model curves is equivalent to Eq.~\ref{oc:eq}, but has a different parametrization to reduce parameter correlations and give more stable fits. The fitted curves use the following parametric function:
\begin{equation}
C(t) = a_0 \sin(a_1 [t-a_2])+a_3 (t-a_2)+a_4,
\end{equation}
where the coefficients are listed in Table~\ref{tab:coefficients}.

\begin{table}[]
    \caption{Coefficients of the model curves in Fig.~\ref{fig:ttvs2}.}
    \centering
    \tiny
    \begin{tabular}{cccccc}
\hline\hline
Curve &&& Coefficients \\
designation \\
\hline
& $a_0$ & $a_1$ & $a_2$ & $a_3$ & $a_4$\\
\hline
1 &13.57 & 0.0049061 & 9410.9 & 0.00833 & $-$2.657 \\
2 &29.18 & 0.0033930 & 9709.9 & 0.04665 & 34.921 \\ 
3 &11.18 &0.0058742 & 9296.77 & $-$0.00016 & $-$8.469 \\
4 &15.85 & 0.0044917 & 9475.23 & 0.01502 & 2.442 \\
5 &19.18 & 0.0040833 & 9548.96 & 0.02380 & 10.170 \\
\hline
\hline
    \end{tabular}
    \label{tab:coefficients}
\end{table}

\label{lastpage}
\end{document}